\documentclass[aps,floatfix]{revtex4}
\usepackage{epsfig}
\usepackage{color}
\usepackage{amssymb}
\usepackage{amsmath}
\usepackage[ngerman,english]{babel}
\usepackage{multirow}
\usepackage{graphicx}

\begin{document}

\title{Correlations and Clustering in Dilute Matter}

\author{G. R\"opke}
\affiliation{Universit\"at Rostock, Institut f\"ur Physik, 18051 Rostock, Germany}
\date{\today}

\begin{abstract}
Nuclear systems are treated within a quantum statistical approach. Correlations and cluster formation are relevant 
for the properties of warm dense matter, but the description is challenging and different approximations are discussed. 
The equation of state, the composition, Bose condensation of bound fermions, the disappearance of bound states 
at increasing density because of Pauli blocking are of relevance for different applications in astrophysics, 
heavy ion collisions, and nuclear structure.
\end{abstract}

\maketitle

\section{Bound states in a fermion systems: The chemical picture}

\subsection{Structure of matter}\label{Sec:1.1}

Our investigation refers to a fundamental property of matter:
We can define elementary constituents (indivisible particles, Greek\,\, $^{^)}\!\!\!\alpha \tau o \mu o \varsigma$) that
 interact. Then, the macroscopic properties are calculated within a quantum statistical approach. 
The interaction may lead to bound states, which can influence the properties in an essential 
manner and new properties may emerge. 
A well known example is the formation of atoms from charged particles, the electrons and
the atomic nuclei. While unbound electrons and nuclei form a plasma which conducts electricity,
the formation of charge-neutral atoms results in  non-conducting dielectric  matter.
Another example is Bose-Einstein condensation which may occur when bosonic bound states are formed in a 
fermionic system. A well-known bosonic bound state is Helium-4, consisting of two electrons, two protons, and two neutrons. 
It shows superfluidity at low temperatures. Nuclear systems where
Bose-Einstein condensation of $\alpha$ particles may occur are discussed below.

Here, we consider nuclear systems consisting of neutrons and protons (and possibly electrons for charge neutrality)  
as constituting "elementary"
particles. As a consequence of the nucleon-nucleon ($N-N$) interaction, bound states, the nuclei, are formed.
Examples are the light nuclei $^2$H (deuteron, $d$),  $^3$H (triton, $t$),  $^3$He (helion, $h$), 
and  $^4$He ($\alpha$ particle). The fundamental interaction is the strong  interaction,
as described by QCD. Because the deduction of the $N-N$ interaction from QCD is not solved at present,
effective interactions are introduced which reproduce measured properties such as binding energies and 
scattering phase shifts. Examples are the Yukawa, Reid, Bonn, Paris, Argonne, etc., interactions, but also the separable (Yamaguchi, PEST, BEST, etc.) parametrization of the $N-N$ interaction are available. 
The optimum form of the effective $N-N$ interaction
is widely discussed in nuclear physics but will not addressed in the present work.

The appearance of bound states is a general feature of the structure of matter. For example, quarks form hadrons,
hadrons form nuclei, nuclei together with electrons form atoms and ions, atoms form molecules.
The new bound states may be considered as the elements of a statistical description of matter,
what is possible as long as the energies considered are small enough so that the internal structure of
the bound states is not excited. A description where the "elementary" particles and the composed bound states are considered 
on the same footing is denoted as chemical picture. In a simple form it is applicable in the low-density 
limit where we have an ideal mixture of the constituents and the composed particles, which are considered as non 
interacting, but accidental reacting so that chemical equilibrium (mass action law) between the different components of the system is established. In the special case of nuclear systems considered here, bound states (nuclei) are expected to appear at low densities 
and low temperatures according to the chemical equilibrium. Such states may occur for expanding matter produced in heavy ion collisions (HIC), but also in astrophysical objects and in excited nuclei.
\begin{table}[ht]
\caption{Bound state formation and dissolution}
{\begin{tabular}{|l|l|l|l|l|l|}
\hline
{\it energy scale}& {\it fermions} & {\it interaction}& {\it bound states} & {\it density effects} & {\it condensed phase} \\
\colrule
$1\dots 10$ meV & electrons, holes  & Coulomb & excitons & screening & electron-hole liquid\\
$1\dots 10$ eV & electrons, nuclei  & Coulomb & ions, atoms & screening & liquid metal\\
$1\dots 10$ MeV & protons, neutrons  & $N-N$ int. & nuclei & Pauli blocking & nuclear matter\\
$0.1\dots 1$ GeV & quarks  & QCD & hadrons & deconfinement & quark-gluon plasma\\
\hline
\end{tabular}}
\label{Tab:1}
\end{table}

\begin{figure}[ht]
 			\includegraphics[width=0.6\linewidth,clip=true]{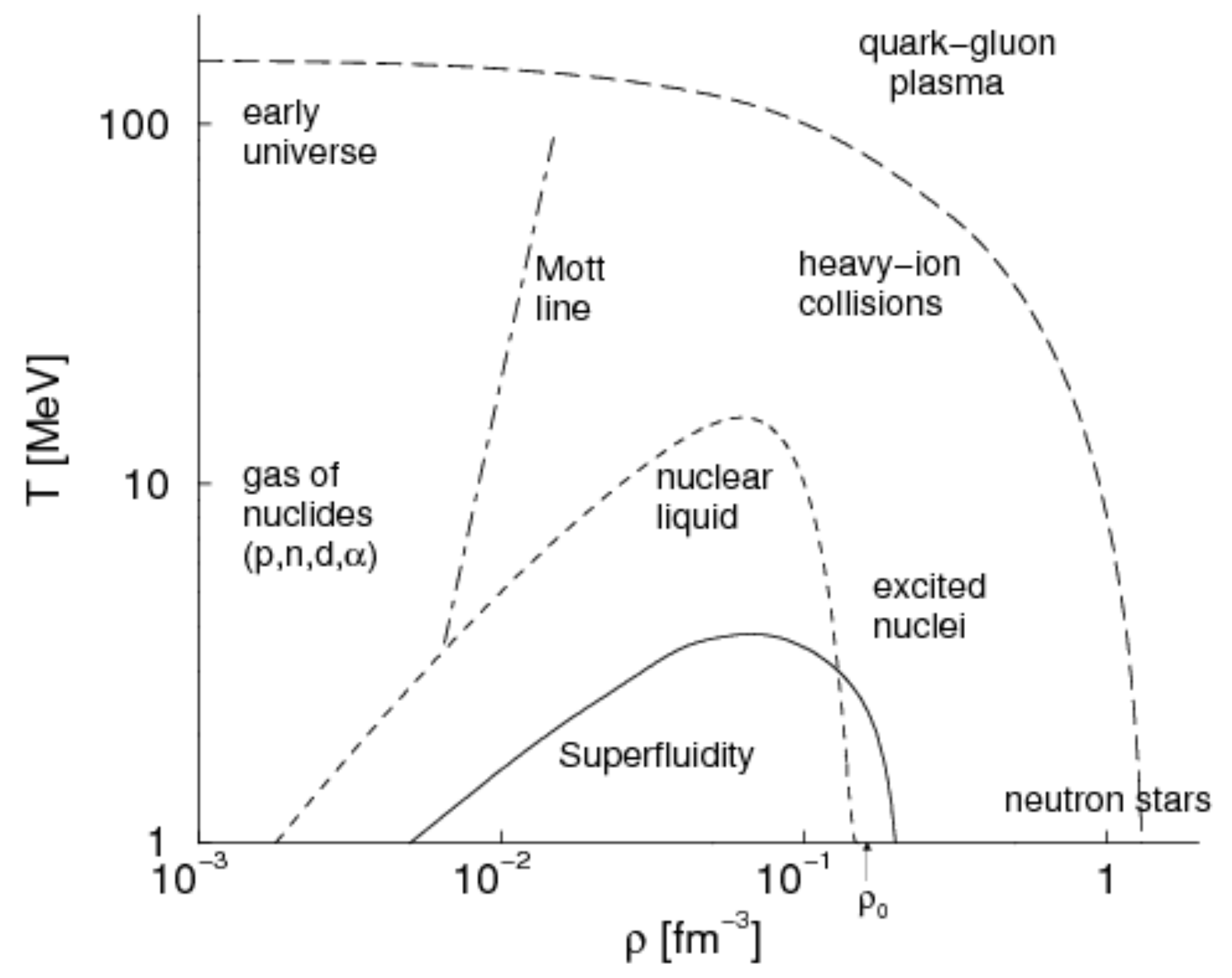}
 			\caption{Phase diagram of symmetric nuclear matter (schematically) \cite{Phasediagram}. Baryon density $n_B=\rho$ with saturation density $n_{\rm sat}=\rho_0$. The Mott line indicates the region where the formation of bound states 
is suppressed because of Pauli blocking.
 }
 			\label{Fig:phasediagram}
 		\end{figure}

A schematic overview of the structure of matter is given in Tab.~\ref{Tab:1}. "Elementary" particles are introduced on different levels, which
prove to be composed particles (bound states) if the scale of energy is changed.
An interesting aspect is the behavior at increasing density. As a general feature, the bound states are dissolved, 
and a phase transition to a condensed phase may occur. Different effects are responsible for the dissolution of bound states:
In Coulomb systems (electrons, ions, and atoms in a plasma; electrons, holes, and excitons in excited semiconductors)
screening of the long-range Coulomb interaction leads to the Mott transition where an atomic insulator goes over to a metallic
conductor; see Ref. \cite{ebe,KKER}. Another effect which leads to the destruction of bound states is the Pauli principle. 
If the constituents of the bound states are fermions, the phase space
which is available to form a bound state is reduced at increasing density (Pauli blocking) so that the formation of bound states 
is suppressed; see Fig. \ref{Fig:phasediagram}. This phenomenon will be discussed in detail below.
It is responsible for the transition from 
a gas of nucleons and nuclei to nuclear matter (nuclear liquid), as it appears at saturation density $n_{\rm sat}\approx 0.16$ fm$^{-3}$.
Also on the level of the quark substructure, at increasing density a deconfinement transition from the hadronic phase 
to a quark-gluon plasma is expected where the hadrons are dissolved.

Starting from a microscopic description of matter, for instance a Hamiltonian with an effective 
$N-N$ interaction, a systematic treatment is given by the quantum statistical approach where correlation functions are evaluated for the equilibrium distribution. 
For infinite matter, we have the temperature $T$ and the chemical potentials $\mu_c$ of the different components $c$. These thermodynamic variables which define the often used grand canonical ensemble are related to the average of the internal energy $U$ and the average particle numbers $N_c=\Omega n_c$, where $\Omega$ denotes a normalization volume (volume of the system) and $n_c$ the density of species $c$. The relations between these variables such as $N_c/\Omega=n_c(T,\mu_{c'})$ are denoted as equations  of state and will be considered in detail in this work. 

Different equations of state are possible. In particular, thermodynamic potentials such as the free energy $F$ as function of $T, n_c$ have the property that all other thermodynamic relations can be obtained by taking
the derivation with respect to the corresponding thermodynamic quantities. In the case considered here, i.e. $\mu_c(T,n_{c'})$,
the free energy is found by integration as $F(T,N,\Omega)=\Omega \int_0^n \mu(T,n') dn'$ (one component). In addition to the thermodynamic properties, it also describes phase transitions when the stability condition $\partial \mu/\partial n|_T\ge 0$ is violated, 
see, e.g., Ref. \cite{R15} for the case of different components. A survey of the phase diagram of symmetric nuclear matter is given in Fig.~\ref{Fig:phasediagram}.

\subsection{Quantum Statistical Calculation of Cluster Abundances in Asymmetric Hot Dense Matter
}\label{Sec:1.2}

Let us start with a brief historical review \cite{R83}. 
The thermodynamic properties of hot nuclear matter are
of interest in connection with the theory of heavy-ion
collisions as well as with astrophysical and cosmological
problems. Of course, the behavior of dense matter under
the conditions of star evolution, the expanding universe
or deep inelastic relativistic ion collisions should be described
by a kinetic theory which gives the detailed nuclear
processes during the time evolution of hot dense matter
in non-equilibrium (see, for instance, Ref.~\cite{bur}). 
However, it is reasonable to compare the result
of these non-equilibrium processes with results for the
thermal equilibrium in the sense of an estimation for the
most probable state the system is likely to attain \cite{kno}. Although we are fully aware of the problem
of using equilibrium results, we suggest that a correctly
formulated theory of thermal equilibrium leads to facts
as the distribution of cluster abundances or possible phase
instabilities which are also of interest for the non-equilibrium
behavior of hot dense matter. For instance, a
thermal model \cite{nag,das} and the concept of a freeze-out baryon density
$n_B$ and a freeze-out temperature $T$ \cite{mek} were
successfully employed in the theory of heavy ion collisions
to determine the production of composite fragments.

\begin{figure}[ht]
 			\includegraphics[width=0.8\linewidth,clip=true]{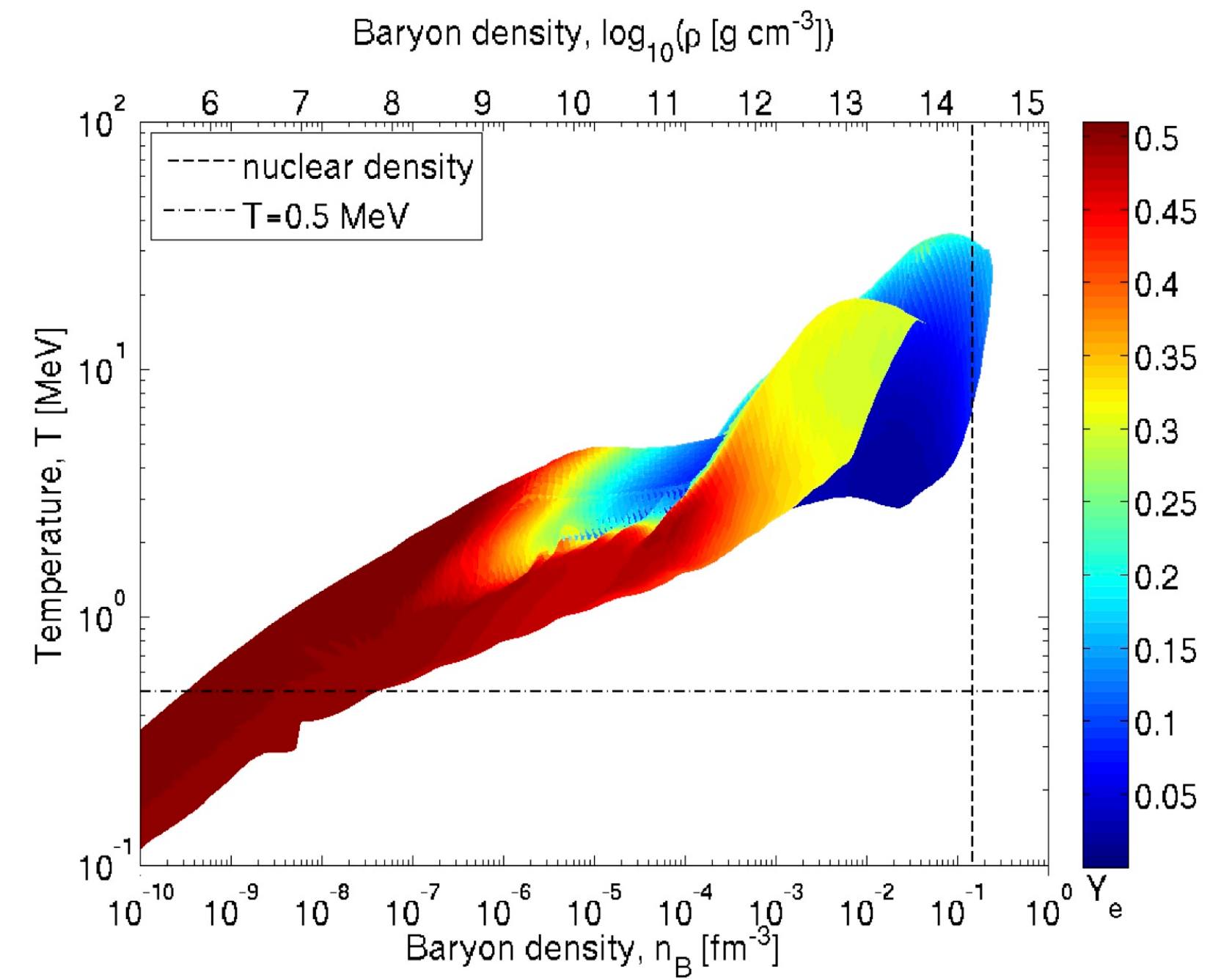}
 			\caption{Exploding supernova. States in the phase diagram which are met during the SN explosion.
			For core collapse SN the region on the phase diagram is slightly reduced. From \cite{Tobias}.
 }
 			\label{Fig:SN}
 		\end{figure}

In the present work, nuclear matter in thermodynamic equilibrium is investigated, 
confined in the volume $\Omega$ at temperature $T$, and consisting
 of $N_n$ neutrons (total neutron density $n^{\rm tot}_n=N_n/\Omega$)
and $N_p$ protons (total proton density $n^{\rm tot}_p=N_p/\Omega$). 
In the thermodynamic limit, the state is given by the parameter set 
$\{T,n^{\rm tot}_n,n^{\rm tot}_p\}$, the dependence on $\Omega$ is trivial.
The subsaturation region will be considered where the baryon density 
$n_B= n^{\rm tot}_n+n^{\rm tot}_p \leq n_{\rm sat}$
(with the saturation density $n_{\rm sat} \approx 0.16$ fm$^{-3}$), the temperature $T \leq 20$ MeV, 
and the proton fraction $Y_p=n^{\rm tot}_p/n_B$ between 0
and 1. This region of warm dense matter is of interest not only for nuclear structure calculations 
and heavy ion collisions 
explored in laboratory experiments \cite{Natowitz,NatowitzEoS}, but also in astrophysical applications, see Ref.~\cite{bro}.
For instance,  core-collapse supernovae at post-bounce stage evolve in this region of the phase space \cite{Tobias,Tobias1}, 
see Fig. \ref{Fig:SN},
and different processes such as  neutrino emission and absorption, which strongly depend on the composition of warm dense matter,
influence the mechanism of core-collapse supernovae. 
		
Let us denote by an ideal nuclide gas an approximation
where nuclear matter is considered as an ideal mixture
of free particles and clusters which move relatively freely
except for occasional nuclear reactions. At equilibrium,
the abundances of clusters are determined by the temperature,
the chemical potentials of the nucleons and the
internal energy of the clusters according to the law of
mass action \cite{mek,nag,das}. No phase transition is
obtained within this simple approximation, and at given
temperature the part of nucleons which is bound in clusters
increases with increasing density.
 
Recently, this simple chemical picture, also called nuclear statistical equilibrium (NSE) \cite{NSE}, has been applied 
for the nuclear matter EoS in the low-density limit. 
Fragmentation as observed in HIC has been treated within a micro-canonical approach
which takes the interaction into account on form of the restriction of available space because of non-overlapping clusters;
for references see \cite{Gross}.

A simple chemical equilibrium of free nuclei is not applicable up to saturation density 
because medium modifications by self-energy shifts and Pauli blocking become relevant.
Alternatively, standard versions \cite{LS,Shen} 
of the nuclear matter equation of state (EoS) considering single nucleons in mean-field (Skyrme or RMF) approximation
as well as a representative heavy nucleus and $\alpha$-particles are available for 
astrophysical simulations such as supernova collapses. 
They have been improved, see Refs.
\cite{SR,Arcones2008,Armen2009,Typel,Gulminelli2010,NSE,Hempel,Furusawa,Connor,shenTeige,Providencia,Avancini,Jaqaman,Vergleich,Hempel2013,Gulminelli2013,Gulminelli2014,NSETabellen,Hempel2014,Constanca,oertel},
elaborating concepts such as the heuristic excluded-volume approach or in-medium nuclear cluster energies 
within the extended Thomas--Fermi approach. These concepts may be applied 
to heavier clusters but are not satisfactory to describe light clusters that require a more fundamental quantum statistical approach.
A generalization of the RMF approach (gRMF) which includes the light nuclei ($A\le 4$) as additional degrees of freedom, where the quasiparticle properties are derived from the quantum statistical approach described below, has been proposed in \cite{Typel}; see also \cite{VT}.
 
A rigorous quantum statistical (QS) approach to the thermodynamic
properties of hot nuclear matter is formulated
within the framework of thermodynamic Green
functions \cite{rop82,rop82a,ropsms}. 
The total number of neutrons $N_n$ and protons $N_p$ are introduced as conserved 
quantities, weak interaction processes leading to $\beta$ equilibrium are not considered.
Starting from the grand-canonical ensemble defined by the temperature $T$ 
and the chemical potentials $\mu_n, \mu_p$ of
neutrons and protons, respectively, the chemical potentials  $\mu_n, \mu_p$ are fixed by the relations
\begin{equation}
\label{eos0}
\frac{1}{\Omega}N_n=n^{\rm tot}_n(T,\mu_n,\mu_p),\qquad \frac{1}{\Omega}N_p=n^{\rm tot}_p(T,\mu_n,\mu_p)
\end{equation}
which are equations of state (EoS) that relate the set of thermodynamic quantities $\{T,\mu_n,\mu_p\}$ 
to $\{T,n^{\rm tot}_n,n^{\rm tot}_p\}$.

The QS approach considers correlation functions and its Fourier transform, the spectral function $S_\tau(1,\omega;T,\mu_n,\mu_p)$.
 The single-nucleon quantum state $|1\rangle$
can be chosen as $1 = \{{\bf p}_1, \sigma_1,\tau_1\}$ which denotes wave number, spin, and isospin, respectively.
A rigorous expression for the nuclear matter EoS is found provided that the spectral function is known,
\begin{equation}
\label{EOS}
  n^{\rm tot}_\tau(T,\mu_n,\mu_p)=\frac{1}{\Omega}\sum_{p_1,\sigma_1} \int \frac{d \omega}{2 \pi} \frac{1}{e^{(\omega-\mu_\tau)/T}+1}
  S_\tau(1,\omega)
\end{equation}
($ \Omega$ is the system volume, $\tau = \{n,p\}$; we take $k_B=1$). 
The spectral function $S_\tau(1,\omega;T,\mu_n,\mu_p)$
is related to the self-energy $\Sigma(1,z)$ for which a systematic approach is possible using diagram techniques, see \cite{FW,AGD}:
\begin{equation}
\label{spectral}
 S_\tau(1,\omega) = \frac{2 {\rm Im}\Sigma(1,\omega-i0)}{ 
(\omega - E(1)- {\rm Re} \Sigma(1,\omega))^2 + 
({\rm Im}\Sigma(1,\omega-i0))^2 }\,;
\end{equation}
$E(1)=\hbar^2p_1^2/2m_1$.

The EoS (\ref{EOS}) relates the total nucleon numbers $N^{\rm tot}_\tau$ or the particle densities $n^{\rm tot}_\tau$ to the 
chemical potentials $\mu_\tau$ of neutrons/protons so that one can switch from the densities to the chemical potentials
characterizing thermodynamic equilibrium of warm dense matter. 
If this EoS is known in some approximation, all other thermodynamic quantities are obtained consistently
after calculating a thermodynamic potential as shown in  Sec. \ref{Sec:1.1}.
In the following sections, different approximations for $\Sigma(1,\omega))$ are discussed.

\subsection{Cluster decomposition of the equation of state and quasiparticle concept}

Within a Hamiltonian approach to the many-particle system, 
the self-energy $\Sigma(1, z)$ may be presented by a series of
diagrams which are constructed from the free nucleon
propagator $G_0^{- 1}(1, z) = z - E(1)$ and the nucleon-nucleon
interaction potential $V(12, 1'2')$.
In order to obtain approximations for the equation of
state (\ref{EOS}) of nuclear matter we can proceed in a number
of different ways \cite{rop82,rop82a}.

The spectral function $S_\tau(1,\omega;T,\mu_n,\mu_p)$ and the corresponding two-point correlation functions (density matrix) are quantities, well-defined in the grand canonical
ensemble characterized by $\{T,\mu_n,\mu_p\}$. 
The self-energy $\Sigma(1,z;T,\mu_n,\mu_p)$ depends, in addition to the single-nucleon quantum state $|1\rangle$, on the complex frequency $z$.
It is calculated at the Matsubara frequencies, the analytical continuation to the $z$ plane must be performed.
Within a perturbative approach it can be represented by Feynman diagrams. A cluster decomposition of the self-energy
with respect to different few-body channels ($c$) is possible \cite{RMS,rop82,rop82a,ropsms}, characterized, for instance, by the nucleon number $A$, as well as spin and isospin variables.

The cluster contributions to the self-energy are derived from an in-medium $A$-particle Schr\"odinger equation which 
describes the propagation of the $A$-particle cluster 
(the Fourier transform of the $A$-particle correlation function gives the corresponding spectral function). 
The  Green function approach provides us with consistent approximations for these few-nucleon propagators. 
In particular, we introduce the quasiparticle concept to describe the propagation of few-nucleon clusters 
(including $A=1$) in warm dense matter if the $A$-particle spectral function $S_A(\omega)$ 
shows a peak structure at the energy $ E_{A,\nu_c}({\bf P};T,\mu_n,\mu_p) $. 
The dispersion relation $E^0_{A,\nu_c}({\bf P})$ of the free nucleon cluster is modified at finite densities.

The Green function approach describes the propagation of a single nucleon by a Dyson equation governed by the 
self-energy, $E_{\tau}({\bf p};T,\mu_n,\mu_p)=E^0_{\tau}({\bf p})+\Delta E_{\tau}^{\rm SE}({\bf p};T,\mu_n,\mu_p)$,
as well as the few-particle states which are obtained from a Bethe-Salpeter equation containing the effective interaction kernel.
Both quantities, the effective interaction kernel and the single-particle self-energy, 
should be approximated consistently.
Approximations which take cluster formation into account have been worked out \cite{ropsms,clustervirial}, 
where within the cluster mean-field (CMF) approximation correlations in the surrounding medium are taken into account.

For the $A$-nucleon cluster, the  in-medium Schr\"odinger equation 
\begin{eqnarray}
&&[E_{\tau_1}({\bf p}_1;T,\mu_n,\mu_p)+\dots + E_{\tau_A}({\bf p}_A;T,\mu_n,\mu_p) 
- E_{A \nu}({\bf P};T,\mu_n,\mu_p)]\psi_{A \nu {\bf P}}(1\dots A)
\nonumber \\ &&
+\sum_{1'\dots A'}\sum_{i<j}[1-n(i;T,\mu_n,\mu_p)- n(j;T,\mu_n,\mu_p)]V(ij,i'j')\nonumber\\
&& \times\prod_{k \neq 
  i,j} \delta_{kk'}\psi_{A \nu {\bf P}}(1'\dots i' \dots j' \dots A')=0
\label{waveA}
\end{eqnarray}
is derived from the Green function approach 
after the effective occupation numbers $n(i;T,\mu_n,\mu_p)$
\begin{equation}
\label{effocc}
n(1) = f_1(1) + \sum^\infty_{\bar A=2} \sum_{\bar \nu \bar {\bf P}} \sum_{2 \dots \bar A}
\bar A \,f_{\bar A, \bar \nu}[E_{\bar A,  \bar \nu}(  \bar {\bf P};T,\mu_n,\mu_p)] |\psi_{\bar A \bar \nu \bar {\bf P}}(1 \dots \bar A)|^2\,,
\end{equation}
 are introduced and exchange terms are neglected.
\begin{equation}
f_{A,Z}(\omega;T,\mu_n,\mu_p)=\frac{1}{ \exp [(\omega - N \mu_n - Z \mu_p)/T]- (-1)^A}
\label{vert}
\end{equation}
is the Bose or Fermi distribution function for even or odd $A$,
respectively, which is depending on $\{T,\mu_n,\mu_p\}$. 

The in-medium Schr\"odinger equation (\ref{waveA}) contains the effects of the medium in the single-nucleon quasiparticle shift 
\begin{equation}
\label{9}
\Delta E_{\tau}^{\rm SE}({\bf p};T,\mu_n,\mu_p) = E_\tau({\bf p};T,\mu_n,\mu_p)-\sqrt{ m^2 c^4+\hbar^2 c^2 p^2} 
+ m c^2 \approx  E_\tau({\bf p};T,\mu_n,\mu_p)-\frac{\hbar^2p^2}{2 m} 
\end{equation}
(nonrelativistic case), as well as in the Pauli blocking terms given by the occupation numbers $n(1;T,\mu_n,\mu_p)$
in the phase space of single-nucleon states $|1 \rangle \equiv |{\bf p}_1,\sigma_1,\tau_1 \rangle$.  
Thus, two effects have to be considered, the quasiparticle
energy shift and the Pauli blocking. 

\begin{figure}[ht]
 			\includegraphics[width=0.6\linewidth,clip=true]{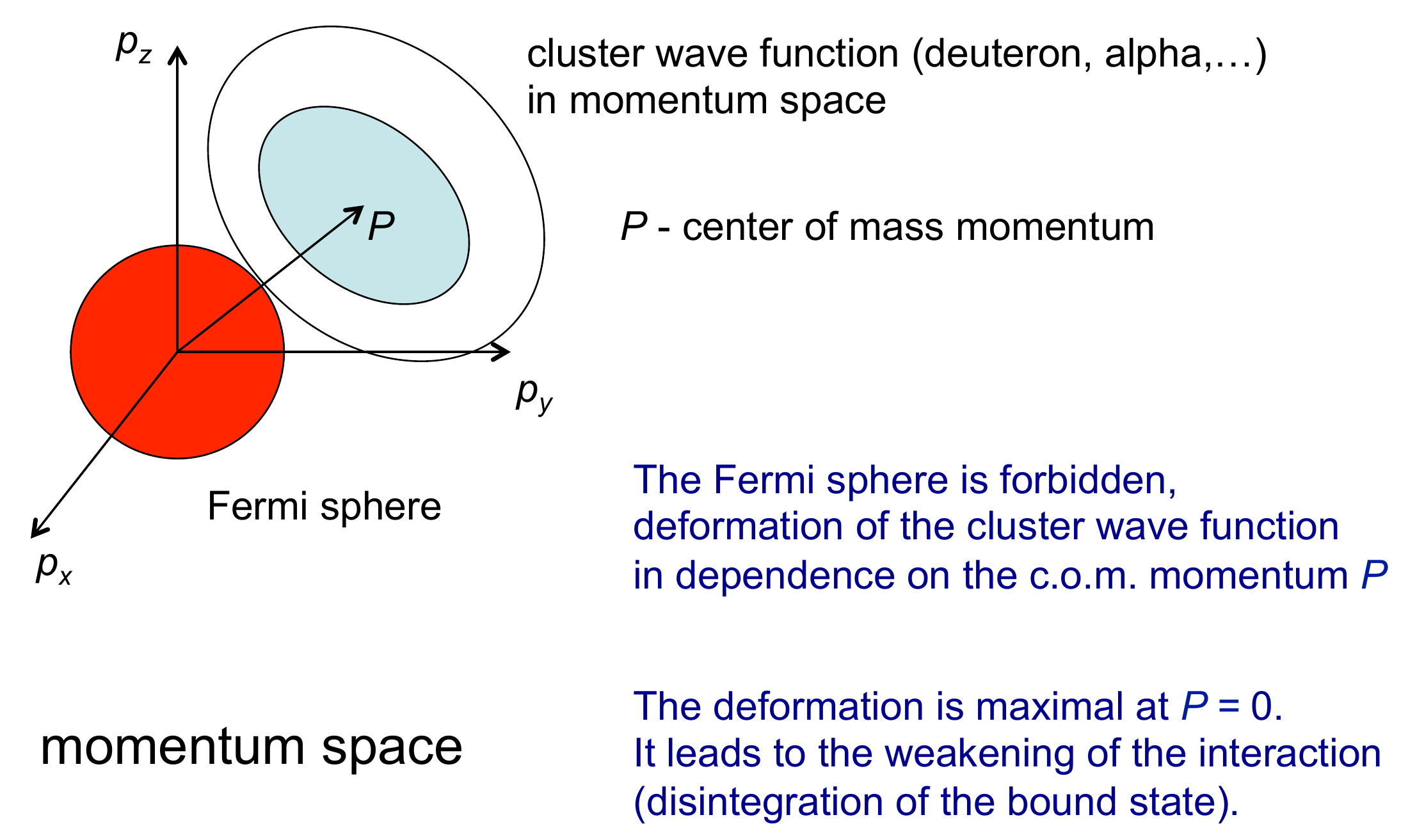}
 			\caption{Bound state wave function in phase space. The Fermi sphere is already occupied by the matter and cannot be used to form a bound state (Pauli principle).
 }
 			\label{Fig:Pauli}
 		\end{figure}

As example, consider the two-nucleon ($A=2$) in-medium Schr\"odinger equation (\ref{waveA})
\begin{eqnarray}
&&[E_{\tau_1}({\bf p}_1;T,\mu_n,\mu_p) + E_{\tau_2}({\bf p}_2;T,\mu_n,\mu_p) 
- E_{2 \nu}({\bf P};T,\mu_n,\mu_p)]\psi_{2 \nu {\bf P}}(12)
\nonumber \\ &&
+\sum_{1'2'}[1-f_1(1)- f_1(2)]V(12,1'2')\psi_{2 \nu {\bf P}}(1'2')=0\,
\label{wave2}
\end{eqnarray}
where the Fermi distribution function is taken for the Pauli blocking term.
As shown in Fig. \ref{Fig:Pauli}, the phase space to form a bound state is reduced owing to the Pauli principle so that
the interaction is effectively reduced and the binding energy is reduced. It should be mentioned that the same equation (\ref{wave2})
 describes the Bose quantum condensation when $E_{\tau_2}({\bf p}_2;T,\mu_n,\mu_p)=\mu_n+\mu_p$, 
as well as the cross-over from BEC to BCS \cite{SRS,Urban,Almpn,Stein,Baldo}.

Using the cluster decomposition of the self-energy which takes into account, in particular, cluster formation,
one obtains
\begin{eqnarray}
\label{eosq}
&&  n^{\rm tot}_n(T,\mu_n,\mu_p)=
 \frac{1 }{ \Omega} \sum_{A,\nu,{\bf P}}N 
f_{A,Z}\left(E_{A,\nu}({\bf P};T,\mu_n,\mu_p)\right), \nonumber\\ 
&&  n^{\rm tot}_p(T,\mu_n,\mu_p)= 
\frac{1 }{ \Omega} \sum_{A,\nu,{\bf P}}Z 
f_{A,Z}\left(E_{A,\nu}({\bf P};T,\mu_n,\mu_p)\right),
\label{quasigas}
\end{eqnarray}
where $\bf P$ denotes the c.o.m. momentum of the cluster (or, for $A=1$, the momentum of the nucleon). 
The internal quantum state $\nu$ contains the proton number $Z$ and neutron number $N=A-Z$ of the cluster.
The integral over $\omega$ is performed within the quasiparticle approach, the $\bf P$-dependent quasiparticle energies $E_{A,\nu}({\bf P};T,\mu_n,\mu_p)$
are depending on the medium characterized by $\{T,\mu_n,\mu_p\}$. These in-medium modifications will be detailed in the following section 
\ref{Sec:1.4}.

\subsection{Different approximations}\label{Sec:1.4}

As it is well known \cite{kad}, the
self-energy occurring in Eq. (\ref{spectral}) may be represented by an
infinite series of irreducible diagrams. We give the physical
ideas how to construct the approximation for the
self-energy used in different approaches. Numerical estimations of
these contributions are given in Sec.\ref{Sec:2}.

In nuclear systems, we are mainly concerned with the strong interaction.
Consequences of the $N-N$ interaction are the 
formation of correlations and bound clusters, but also self-energy shifts 
and the Pauli exclusion principle. A free nucleon feels a
mean Hartree field due to the surrounding nuclear matter
consisting of free nucleons and clusters, which is modified
by the Pauli exclusion principle ("Fermi hole") so that
the Hartree-Fock single particle energy shift $\Delta^{\rm HF}(1)$  is obtained,
\begin{equation}
\label{3}
\Sigma^{\rm HF}(1, z)=\Delta^{\rm HF}(1)=\sum_2 [V (12, 12) - V (12. 21)] f _{1,\tau_2}(2),
\end{equation}
where $f _{1,\tau_1}(1) = [\exp (E(1) - \mu_{\tau_1})/T + 1]^{-1}$ is the Fermi distribution
function.

Similarly, a bound state (cluster) of $A$
nucleons with total momentum $P$ and internal quantum
state $\nu$ (including the proton number $Z$) feels a self-energy
shift $\Delta E^{\rm SE}_{A\nu P}$ due to the surrounding matter. In addition,
the binding energy is lowered by $\Delta E^{\rm Pauli}_{A\nu P}$ as a consequence
of the Pauli exclusion principle, because phase
space is occupied by the surrounding correlated matter
and is not available to form a bound state. The lowering
of the binding energy leads to the destruction of bound
states in dense matter.

Notice that also the other forces (weak, Coulomb, gravitation) are of relevance describing
matter in astrophysics of compact objects not considered here. $\beta$ equilibrium
is not achieved in HIC because the time scales are short.
Coulomb interaction contributions are
of importance especially for heavy clusters because of its
long range character. In the region of thermodynamic instability of nuclear matter,
it is essential for the formation of pasta structures. The QS treatment of the Coulomb interaction
is well elaborated, see Ref. \cite{KKER}.

Coming back to the strong interaction,
we consider Eq. (\ref{eosq}) together with (\ref{waveA}) as the basic result for the EoS.
They are treated at different levels of sophistication as will be explained from a very general point of view,
see Tab. \ref{Tab:2}. This allows us to compare different approximations presently 
used to describe properties of dense matter. The consistent treatment of different effects is 
clearly demonstrated.

\begin{table}[ht]
\caption{Different approximations.}
{\begin{tabular}{|l|l|}
\hline
{\it low density limit }& {\it high density}\\
 (1)  & (1 medium)\qquad {\bf medium effects}\\
\colrule
{\it Ideal Fermi gas: }& {\it Quasiparticle quantum liquid:}\\
neutrons, protons  &mean-field approximation\\
(electrons, neutrinos,\dots) &Skyrme, Gogny, RMF \\
 \colrule
(2) {\bf bound state formation} & (2 medium)\\
 \colrule
{\it Nuclear statistical equilibrium:} & {\it Chemical equilibrium
of quasiparticle clusters:}\\
ideal mixture of all bound states & medium modified bound state energies\\
 chemical equilibrium, mass action law& self-energy and Pauli blocking\\
 \colrule
(3) {\bf continuum contributions} & (3 medium)\\
 \colrule
{\it Second virial coefficient: }& {\it Generalized Beth-Uhlenbeck
formula:}\\
account of continuum correlations ($A=2$)& medium modified binding energies,\\
scattering phase shifts, Beth-Uhlenbeck Eq.
  &medium modified scattering phase shifts\\
 \colrule
(4) {\bf chemical \& physical picture} & (4 medium)\\
 \colrule
{\it Cluster virial approach:}  & {\it Correlated medium:}\\
all bound states (clusters) & phase space occupation by all bound states\\
scattering phase shifts of all pairs & in-medium correlations, quantum condensates \\
\hline
\end{tabular}}
\label{Tab:2}
\end{table}

(1) The simplest approximation is obtained if cluster formation and mean-field  effects, 
i.e. any effects of interaction, are neglected (zero self-energy). We obtain the 
ideal Fermi gas consisting of 
protons, neutrons, (electrons, neutrinos,\dots) well known from text books.
In Eq. (\ref{eosq}), we have only $A=1$ and the free fermion energies $E_\tau^0({\bf p})$.

This simple approximation can be improved in two directions:

(1 medium) At one hand by including mean-field effects when going to higher densities. 
We obtain a quasiparticle quantum liquid. 
Expanding $\Sigma(1, z)$ in a power series with respect to
$V(12, 1'2')$, in lowest order we obtain the Hartree-Fock
approximation (\ref{3}). 
Investigations of hot nuclear matter
within the frame of such a Hartree-Fock approximation
have been performed, see, for instance, Refs. \cite{lam,bar,fri}. A region of thermodynamic
instability was found, and the occurrence of a first
order phase transition below a critical point $T_c\approx 20$ MeV
has been discussed there. No clusters can be found within
a Hartree-Fock approximation.

(2) On the other hand,
expanding $\Sigma(1, z)$ with respect to the nucleon density,
i. e. for $n_c \Lambda_c^3/2 \ll 1$ with $\Lambda_c = (2\pi \hbar^2/m_cT)^{1/2}$ being the
thermal wavelength, in the lowest order the Beth-Uhlenbeck
formula is obtained for the second virial coefficient. For
hot dense matter, this approximation has been discussed,
e. g., in Refs. \cite{mek,jen}. The Beth-Uhlenbeck formula takes
into account the formation of two-particle clusters. 
However, no in-medium corrections for these clusters are
obtained.

To obtain the ideal nuclide gas approximation (NSE), all
bound states should be taken into account on the same
footing. This may be done using a cluster decomposition
for $\Sigma(1, z)$ using the t-matrices of the isolated $A$-particle
system \cite{rop82,rop83}. This approximation is the basis of
the ordinary thermal model \cite{mek,maz,ele,nag,das} 
widely used in describing the occurrence
of clusters in hot dense matter.

(2 medium) In the high density region, the ideal nuclide gas
approximation becomes not applicable because of density
corrections due to the interaction of the clusters with the
surrounding matter. A systematic approach to the in-medium
corrections for the free particle and bound state
energies as well as the wave functions can be given
within the framework of many-body theory. A self-consistent
ladder Hartree-Fock approximation \cite{rop83} has been applied to the quantum statistical calculation
of the abundances of deuterons, tritons and alpha
particles \cite{mun} and to the description
of a nuclear matter phase transition \cite{sch82,sch83}.
As discussed in these papers, the in-medium corrections
to the energy and wave functions of the clusters lead to
an interesting effect: At high densities, bound states are
destroyed because of the Pauli quenching. Beyond a certain
density (Mott density $n_A^{\rm Mott}(T)$) the abundances of
the corresponding clusters decrease, and in the high
density limit all bound states are dissolved so that a degenerate
Fermi liquid of quasiparticles remains.

(3) In Eq. (\ref{eosq}), the summation refers to the mass number $A$, 
the internal quantum number $\nu$ and the c. o. m. momentum $\bf P$.
The internal quantum number $\nu$ includes in addition to the 
proton number $Z$ also excited bound states with medium
dependent energies $E_{A,\nu}({\bf P}; T,\mu_n,\mu_p)$
as well as continuum states. In the two-particle case, these
continuum contributions are expressed by the scattering phase shifts 
as function of the energy of relative motion. Only after including the 
continuum contributions, the correct expression for a virial expansion 
is obtained. The corresponding relation is known as Beth-Uhlenbeck 
equation. 

(3 medium) To pass to higher densities, in addition to  medium modified 
bound state quasiparticle energies also the medium modified
scattering phase shifts have to be calculated. For $A=2$, the generalized 
Beth-Uhlenbeck formula \cite{SRS} is obtained. As a peculiarity,
when introducing the quasiparticle description one has to subtract 
contributions of the continuum scattering phase shifts to avoid double counting.

(4) This concept of inclusion of scattering state contributions is
given in the cluster Beth-Uhlenbeck approach where arbitrary 
mass numbers $A$ are considered \cite{Debrecen}. Only 
estimates for the continuum correlations are known at present \cite{R15},
and the in-medium modifications of the corresponding cluster-virial
coefficients are relevant for the composition and the EoS.

(4 medium) A consistent description of the medium effects should also take into account
correlations in the medium. Extending the Hartree-Fock approximation to arbitrary clusters, the CMF approximation has been derived \cite{ropsms,R15}.
With respect to the Pauli blocking (\ref{waveA}), the phase space occupation numbers
$ n(1;T,\mu_n,\mu_p) $ are not given by the free particle fermion distribution $f_{\tau_1}(p_1)$ but the effective occupation numbers (\ref{effocc}) 
which considers also the phase space occupation owing to correlations and to the formation of clusters. An estimate is given in Ref. \cite{R15}. 
For a more systematic approach, the self-consistent determination of in-medium correlations and cluster formation has to be treated which is not solved until now.
 For instance, the unified description of $\alpha$ matter and nuclear matter is not known at present.

\section{Model calculations for in-medium corrections of cluster
energy values}
\label{Sec:2}

Model calculations are performed to describe correlations and cluster formation in nuclear systems.
We have to go beyond single-particle descriptions which have ben proven to be very successful
not only in nuclear structure calculations where the shell model has been worked out, but also
in the thermodynamic properties of nuclear matter where mean-field approaches are very popular.
Nuclear systems at low density and low excitation energy are dominated by correlations and cluster formation,
if the kinetic energy characterized by the temperature or the Fermi energy is small compared with the 
potential energy.

Starting from the interaction potential $V(12,1'2')$ which
contains the long range Coulomb interaction as well as the
short range nucleon-nucleon interaction, {\it ab initio} calculations
of correlations, in particular structure of bound states ($A \nu P$) are rather involved. 
In addition, in contrast to the fundamental Coulomb interaction which is known,
the $N-N$ interaction is introduced in an empirical way, fitted to measured properties of nuclear systems.
Consequently, QS calculations for nuclear systems should implement as much as possible measured data,
for instance the empirical binding energies of isolated nuclei \cite{Audi}.

We focus on the contributions of the $N-N$ interaction to the structure of nuclear systems. 
The Coulomb interaction is treated using approaches known from plasma physics \cite{ebe,KKER}; see Sec. \ref{Sec:2.1}.
It has to be taken into account for nuclear structure and nuclear reactions.
In particular, in context with the EoS and the composition of nuclear matter, 
it becomes relevant for heavy nuclei and pasta-like structures.

\subsection{Single-quasiparticle approximation}
\label{Sec:2.1}
Within the QS approach, the influence of the medium is given by the self-energy $\Sigma(1,z)$.
It fixes the spectral function, Eq. (\ref{spectral}) which then allows to calculate the EoS (\ref{eosq}).
For the self-energy, a systematic approach is possible using diagram techniques, see \cite{FW,AGD}.
The ideal Fermi gas is a trivial approximation of the EoS, Eq. (\ref{eosq}), neglecting any self-energy 
contribution in Eq. (\ref{spectral}) so that only $A=1$ with the free dispersion relation 
$E(1)=\hbar^2p_1^2/2m_1$ remains.

We discuss the approximation (1, medium) of Tab. \ref{Tab:2}, considering single-nucleon quasiparticle states
in dense nuclear matter.
In lowest order with respect to the interaction, neglecting  correlations in the surrounding matter
the Hartree-Fock (HF) result (\ref{3}) is obtained.
The approximation $\Sigma^{\rm HF}(1,z)= \Delta^{\rm HF}(1)$ is real and not depending on the frequency $z$.
This is in correspondence to the Kramers-Kronig relation that any frequency dependence of $\Sigma(1,z)$ would 
produce an imaginary contribution Im $\Sigma(1,z)$. Because Im $\Sigma^{\rm HF}(1,z)=0$, the spectral function 
(\ref{spectral}) depends on $\omega$ as
\begin{equation}
 S_\tau(1, \omega)= 2 \pi \delta[\omega-E(1)-\Delta^{\rm HF}(1)].
\end{equation}
In HF approximation, we have a sharp quasiparticle with the shifted energy $E^{\rm qu, HF}(1)=E(1)+\Delta^{\rm HF}(1)$. This modification 
of the single-particle energy $E(1)$ is denoted as mean field.

In higher order with respect to the interaction, contributions are expected for Im $\Sigma(1,z)$ describing collisions in the system.
The $\delta$-like spectral function obtained from the quasiparticle pole $\omega= E^{\rm qu}(1)$, 
with the self-consistent solution of
\begin{equation}
\label{quasi}
  E^{\rm qu}(1) =E(1)+{\rm Re}\,\Sigma(1,\omega= E^{\rm qu}(1)),
\end{equation}
 is broadened by Im $\Sigma(1,z)$ which describes the 
finite life-time of the quasi-particle.

The Hartree-Fock approximation has been improved taking into account two-particle correlations
 within the Brueckner approach, see  \cite{SRS,Wolter}. From the self-energy the spectral
function is calculated (\ref{spectral}). The position of the peak, i.e. the self-consistent solution of Eq. (\ref{quasi}) 
gives the quasi-particle self energy $E_\tau({\bf p}; T,\mu_n, \mu_p)$. However, the peak structure  
is not always clearly seen from the spectral function,   for instance at low densities, where bound states
appear.   Near the saturation density phenomenological values for 
different properties such as the saturation density, binding energy, compressibility are known which 
are not correctly obtained within the Brueckner theory using a simple form of $N-N$ interaction.

Instead of microscopic calculations of the quasiparticle energies, based on a semi-empirical $N-N$ interaction,
we can also directly parametrize the quasiparticle energies using the known properties of nuclear matter.
A simple parametrization is given by Skyrme; see \cite{Skyrme,vau}. 
The Hartree-Fock shift (\ref{3}) is estimated using, for instance, a
 zero range effective interaction $V(n_B) \delta({\bf r}_1-{\bf r}_2)$
 so that $\Delta E_{\tau}^{\rm SE}(1)$ is independent
of momentum and temperature. 

A more advanced parametrization is the
relativistic mean-field theory (RMF) where the quasiparticle energies are parametrized by a scalar $S(T,n_B,Y_p)$ and
a vector potential $V_\tau(T,n_B,Y_p)$; see Refs. \cite{Typel1999,Typel2005}, adapted to empirical data for nuclear systems:
\begin{equation}
\label{singqu}
E_\tau({\bf p};T,n_B,Y_p)=\sqrt{[m_\tau c^2-S(T,n_B,Y_p)]^2+\hbar^2c^2p^2}+ V_\tau(T,n_B,Y_p)-m_\tau c^2.
\end{equation}
In the limit $p \to 0$, the quasiparticle dispersion relation leads to the effective mass approximation
\begin{equation}
\label{effm}
 E_\tau({\bf p};T,n_B,Y_p) \approx \Delta E_\tau^{\rm SE}(0;T,n_B,Y_p)+\frac{\hbar^2}{2 m^*_\tau (T,n_B,Y_p)}p^2
\end{equation}
with the shift 
$\Delta E_\tau^{\rm SE}(0)=-S(T,n_B,Y_p)+V_\tau(T,n_B,Y_p)$ and the effective mass ratio 
$m^*_\tau/m_\tau =1-S(T,n_B,Y_p)/(m_\tau c^2)$. Explicit expressions for $S(T,n_B,Y_p)$ and $V_\tau(T,n_B,Y_p)$
in form of Pad{\'e} approximations which are suitable for numerical applications are found in \cite{clustervirial,R15,Typel}.

As a mean-field approach, in the low-density limit a linear dependence $S,V_\tau \propto n_B$ is assumed.
This may be a good choice for the shift of the quasiparticle peak as solution of Eq. (\ref{quasi}). 
The EoS (\ref{eosq}) obtained taking only $A=1$ in quasiparticle approximation gives already a reasonable description near 
the saturation density. In addition, a region of thermodynamic instability $\partial \mu/\partial n_B <0$
is obtained below a critical temperature (symmetric matter) $T_{\rm crit}^{\rm mf} \approx 13.72$ MeV \cite{Typel} indicating
the occurrence of a phase transition.
No cluster formation is described in the single-quasiparticle approximation. Therefore, the RMF approximation is not sufficient in the low-density, low-temperature region where bound states occur. 

The concept to parametrize the single-nucleon quasiparticle energies is related to the density functional theory
used in condensed matter theory. A problem is the exact treatment of correlations. Part of the interaction is already
implemented in the parametrization of the energy density functional. If the correlations are considered separately
one has to avoid double counting. We come back to this issue below.

\subsection{Cluster-quasiparticle approximation}

To obtain the formation of clusters
and correlations, the term ${\rm Im} \, \Sigma$ in the spectral function (\ref{spectral}) has to
be analyzed. A cluster decomposition of the self-energy gives the contribution of the $A$-nucleon correlation, 
in particular the bound states. Neglecting all medium effects in the few-nucleon wave equation (\ref{waveA}),
the ideal nuclide gas approximation (NSE) for the EoS (\ref{eosq}) is obtained. Instead solving the isolated
$A$-nucleon Schr\"odinger equation with an appropriate $N-N$ interaction to obtain the energy eigenvalues $E_{A,\nu}^{(0)}({\bf P})$,
we can directly use the empirical values for the masses of nuclei as given, e.g., by Ref. \cite{Audi}.

To describe medium effects consistently with the single-quasiparticle approximation, the few-nucleon wave equation (\ref{waveA})
has to be solved. Within a perturbative approach, the cluster-quasiparticle shift has two contributions ($A>1$),
\begin{equation}
 E_{A,\nu}({\bf P};T,\mu_n,\mu_p)=E_{A,\nu}^{(0)}({\bf P})+\Delta E^{\rm SE}_{A,\nu}({\bf P};T,\mu_n,\mu_p)
+\Delta E^{\rm Pauli}_{A,\nu}({\bf P};T,\mu_n,\mu_p).
\end{equation}
In addition to the single-quasiparticle energy shift 
$\langle \psi_{A\nu {\bf P}}|\sum_i \Delta E_{\tau_i}({\bf p}_i)|\psi_{A\nu {\bf P}} \rangle$, 
the Pauli blocking $-\langle \psi_{A\nu {\bf P}}|\sum_{i \neq j} n(i) V(ij,i'j') |\psi_{A\nu {\bf P}} \rangle$ 
must be considered to obtain the 
approximation (2, medium) of Tab. \ref{Tab:2}.

Within a simple estimate for the medium-modified cluster-quasiparticle energies,
the cluster self-energy shift is easily calculated in the rigid shift approximation 
\begin{equation}
 \Delta E^{\rm SE}_{A\nu P}=\sum_i^A \Delta E_{\tau_i}^{\rm SE}(0;T,n_B,Y_p)
\end{equation}
neglecting the effective mass corrections. To improve it, the effective mass approximation (\ref{effm}) can be performed using
empirical values for $m_\tau^*$, see \cite{R15}. 

\begin{figure}[ht]
 			\includegraphics[width=0.7\linewidth,clip=true]{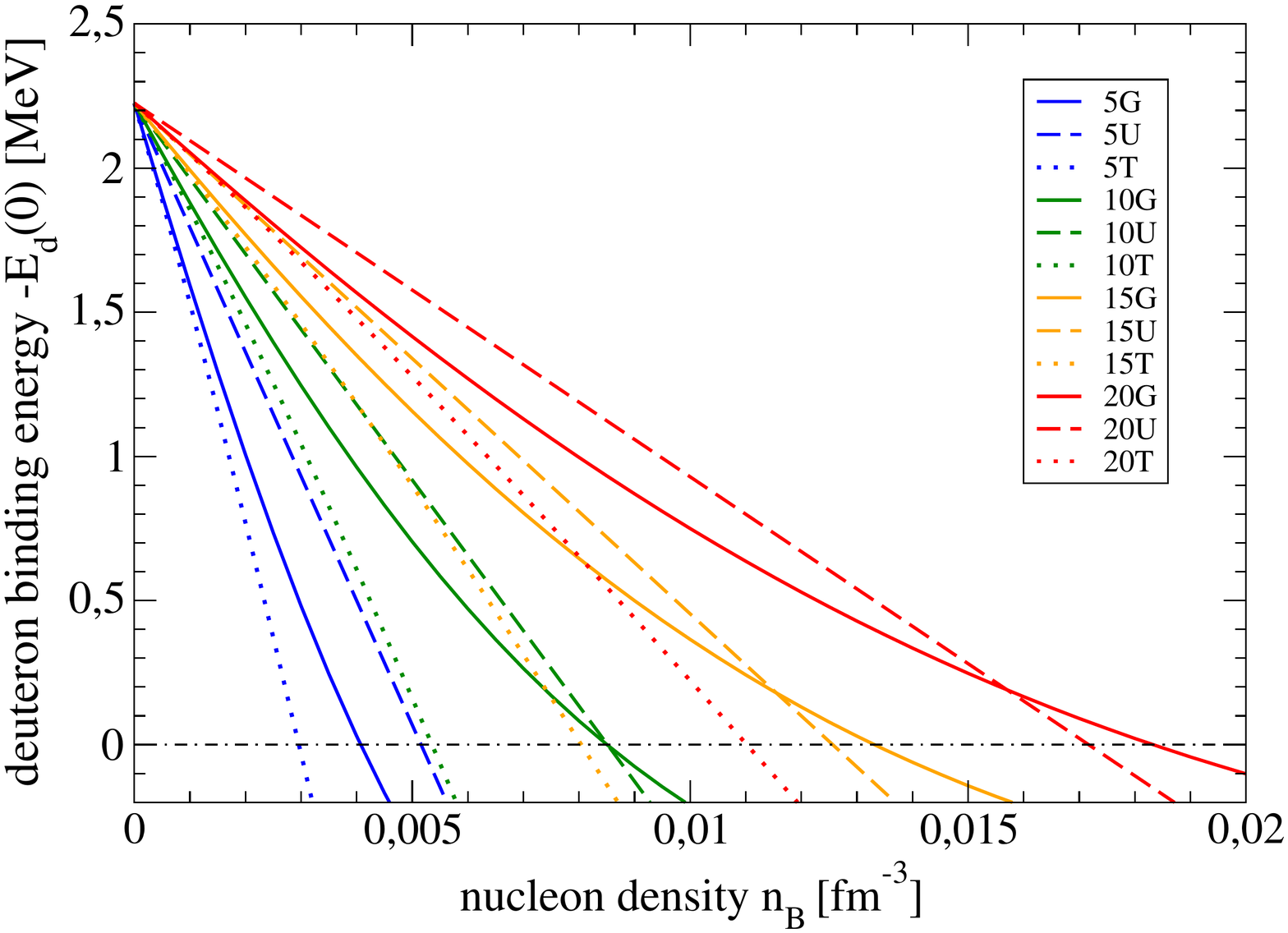}
 			\caption{Shift of the deuteron binding energy because Pauli blocking. Temperatures $T=5, 10, 15, 20$ MeV (from left to right), $P=0$. 
The rigorous result \cite{R15,R2011} (G, full) is compared with the simple estimation (\ref{Pd}) (U, broken) 
and the expression given in Ref. \cite{Typel} (T, dashed).
 }
 			\label{Fig:Udo1}
 		\end{figure}

In the Pauli quenching term 
\begin{equation}
\label{DeltaP}
\Delta E^{\rm Pauli}_{A\nu P}=-\sum_{1\dots A}  \sum_{1'\dots A'}
 \psi_{A\nu P}^{0*}(1 \dots A) \psi_{A\nu P}^{0}(1' \dots A') {\sum_{ij}^A}{'}
f_{1,\tau_i}({\bf p}_i-{\bf P}/A)V(ij,i'j')\delta_{kk'}
\end{equation}
the interaction
potential $V(12, 1'2')$ may be approximately eliminated if the cluster wave
function is represented by the antisymmetrized product
of single particle wave functions $\phi_n(i) $ \cite{R83,rop83}.
In the case $A=2$ this can be done rigorously. After separating the c. o. m. motion $\bf P$,
the relative motion (relative momentum $\bf p$) of the free deuteron obeys a Schr\"odinger equation
$ \sum_{{\bf p}'} V({\bf p},{\bf p}')\psi_d^{(0)}({\bf p}')=(E_d^{(0)}-\hbar^2p^2/m)\psi_d^{(0)}({\bf p})$
so that
\begin{equation}
\label{DeltaP2}
 \Delta E^{\rm Pauli}_{d, P}=\int \frac{d^3 p}{(2 \pi)^3} \left[f_n\!\left(\frac{\bf P}{2}+{\bf p}\right)
+f_p\!\left(\frac{\bf P}{2}-{\bf p}\right)\right]\left(-E_d^{(0)}+\hbar^2p^2/m \right)|\psi_d^{(0)}({\bf p})|^2 .
\end{equation}
With the bound state energy $E_d^{(0)}=-2.225$ MeV, the Gaussian wave function (which can be replaced by
a better wave function corresponding to the empirical density distribution) 
$\psi_d^{(0)}({\bf p})=(8^{1/2} \pi^{3/4}/k^{3/2}_d) \exp(-p^2/2 k^{2}_d)$ the evaluation of Eq. (\ref{DeltaP2}) 
is simple and straight forward.
The parameter $k_d = 0.312$ fm$^{-1}$ is fixed by the nucleonic point rms radius \cite{R2009}. 
In the low-density limit
where $f_\tau(p)=n^{\rm tot}_\tau \Lambda^3/2 \exp(-\hbar^2p^2/2 m T)$ with the thermal wave length 
$\Lambda = (2 \pi \hbar^2/m T)^{1/2}$ the result reads  
\begin{equation}
\label{Pd}
 \Delta E^{\rm Pauli}_{d, P}=\frac{n_B \Lambda^3 T^{3/2}}{2 (2 \epsilon_d+T)^{3/2}} e^{- X_d^2 \frac{\epsilon_d+T}{2 \epsilon_d+T}}
\left(-E_d^{(0)}+\frac{3 \epsilon_d T}{2 \epsilon_d+T}+X_d^2 \frac{2 \epsilon_d^2 T}{(2 \epsilon_d+T)^2}\right)
\end{equation}
with $X_d^2=\hbar^2 P^2/8 m T$ and $\epsilon_d=\hbar^2 k_d^2/2 m =2.02$ MeV. In addition to the linear dependence on the nucleon density 
$n_B$ as a consequence of the perturbation approach, the strong variation of the Pauli blocking shift with $P$ and $T$ is remarkable, in contrast with the single-nucleon mean-field shifts. For a rigorous result, 
the empirical form factor $|\psi_d^{(0)}({\bf p})|^2$ has to be used instead of a Gaussian form factor, fitted to the rms radius.
A more sophisticated evaluation of $\Delta E^{\rm Pauli}_{d, P}$ is found in Refs. \cite{R2011,R15} where also Pad{\'e} approximations for the dependence on $\{{\bf P}, T,n_B,Y_p\}$ are given, see Fig. \ref{Fig:Udo1}.
(Note that the shift given in Ref. \cite{Typel}, Eq. (72), contains an empirical factor to suppress the abundance at higher densities.)

For the evaluation of the Pauli quenching term for the other light elements $A \le 4$, i.e. $t, h, \alpha$, assumptions for the 
interaction potential are necessary. Even for the product ansatz, the binding energy $E^{(0)}_{A\nu P}$ 
is not given by the sum of the single nucleon state energies $E_n$ so that shell model calculations have to be used,
\begin{equation}
\label{29}
 \Delta E^{\rm Pauli}_{A\nu P}=-\sum_{n}^{\rm occ.} \sum_1 |\phi_n(1)|^2 
(E_n-E_1)f_{1,\tau_1}({\bf p}_1-{\bf P}/A).
\end{equation}
Estimates based on the harmonic oscillator model are given in  \cite{R83,rop83}.
A more sophisticated calculation where the c.o.m. motion is eliminated, 
has been performed using a separable potential in Ref. \cite{R2011,R15}. 
Pad{\'e} approximations for $\Delta E^{\rm Pauli}_{A 0 P}({\bf P}, T,n_B,Y_p)$ are given there 
which are suited for the evaluation of the EoS (\ref{eosq}) and the corresponding thermodynamic potentials. 
A more simple parametrization of the quasiparticle shifts of the light elements has been considered in Ref. \cite{Typel} where also 
various thermodynamic functions have been presented.
The quasiparticle shifts $ \Delta E_{A 0 P} $ of light clusters, depending on $\{ {\bf P}, T,n_n, n_p\}$,
have been investigated recently \cite{R15,R2011,R2009}. In-medium bound state energies are calculated for 
the light clusters ($d,t,h, \alpha$), and fit formulae are presented there.

Clusters with $5 \le A \le 11$ (light metals) are weakly bound and therefore very sensitive to the medium effects,
similar to the deuterons.
These elements are weakly bound (note that $^8$Be is unbound) and experience a strong influence of the medium. 
A strong depletion because of the Pauli blocking is expected. Estimates of $\Delta E^{\rm Pauli}_{A 0 P}$ for $4 < A \le 16$
are given in Refs. \cite{ropsms,rop83,R83,Debrecen}.

Different approaches are used to calculate the effect of medium modification for large nuclei. The heavy clusters $A \ge 12$ are usually considered as droplets of a dense phase with $n_B \approx n_{\rm sat}$.  As a semiempirical treatment of medium effects, in particular Pauli blocking, the excluded-volume model \cite{Hempel,Hempeleos} may be introduced.
As an alternative, local density approaches such as the Thomas-Fermi model can be applied to calculate the modification of the 
cluster in a dense medium \cite{LS,Shen,Furusawa,Gulminelli2013,Gulminelli2014,Avancini,Constanca} which
 provide us with a microscopic treatment of large nuclei in a dense medium.

We give here an approach \cite{ropsms,rop83,R83} which is closely related to the treatment of light clusters.
The total energy shift of a cluster ($A \nu P$)
\begin{eqnarray}
\label{34}
&& \Delta E_{A \nu P}(T,n_B,Y_p)=\Delta E^{\rm SE}_{A\nu P}+\Delta E^{\rm Pauli}_{A\nu P}
=\sum_\alpha^{\rm occ.} \sum_{22'} \phi_\alpha^*(2)\phi_\alpha(2') \\
&& \times \left\{\sum_{11'} f_{\tau_1}\left({\bf p}_1-\frac{\bf P}{A}\right) \left[\delta_{11'} \delta_{22'}
-\sum_{\alpha'}^{\rm occ.} \phi_{\alpha'}^*(1)\phi_{\alpha'}(1')\right] [V(12,1'2')-V(12,2'1')]\right\}\nonumber
\end{eqnarray}
has a simple structure if within a homogeneous Fermi gas
model the wave functions $\phi_\alpha(i)$ are taken as momentum
eigenfunctions. At not too high temperatures, a strong compensation
between the self-energy and the Pauli quenching
term results. Thus large clusters remain nearly unshifted.
Applying the homogeneous Fermi gas model to a cluster
with densities $n_\tau^A(r)$ for the protons and neutrons, respectively,
we obtain
\begin{eqnarray}
\label{35}
\Delta E_{A \nu P}(n_B)&=&\sum_{\tau=n,p}\int d^3r \Delta E_{\tau}^{\rm SE}(n^A_n(r),n^A_p(r)) n_\tau\\
&& \times\int^\infty_{\Lambda_\tau p_F(n_\tau^A(r))}\frac{y dy}{2 \pi x_\tau}\left[e^{-(y-x_\tau)^2/4 \pi}-e^{-(y+x_\tau)^2/4 \pi}\right]\nonumber
\end{eqnarray}
with
\begin{eqnarray}
\label{35a}
&& p_F(n_\tau)= (3 \pi^2 n_\tau)^{1/3},\qquad x_\tau = m_\tau v \Lambda_\tau/\hbar \approx P \Lambda/A.
\end{eqnarray}

An analysis of experimental data was performed to fit the following symmetrical density
distribution function (see, e.g., Ref. \cite{burov})
\begin{equation}
 n^A(r)=\frac{3 A}{4 \pi R^3} \frac{1}{1+(\pi b/R)^2}\left[\frac{1}{1+e^{(r-R)/b}}+\frac{1}{1+e^{(-r-R)/b}}\right].
\end{equation}
Appropriate values for $A > 16$ are $R = 1.05$ fm $A^{1/3}$,
$b = 0.57$ fm. Furthermore we take $n_p^A (r) = n_B^A (r)Z/A$.

\subsection{Mott points and Mott momenta}
\label{sec:Mott}
Of special interest are the binding energies 
\begin{equation}
\label{Ebind}
B^{\rm bind}_{A,\nu}({\bf P};T,n_B,Y_p)
=-[ E_{A,\nu}({\bf P};T,n_B,Y_p)-E^{\rm cont}_{A,\nu}({\bf P};T,n_B,Y_p)]
 \end{equation}
with 
$E^{\rm cont}_{A,\nu}({\bf P})=N E_n({\bf P}/A)+Z E_p({\bf P}/A)$,
that indicate the energy difference between the bound state and the continuum of free (scattering) states at the same total momentum ${\bf P}$. 
This binding energy determines the yield of the different nuclei 
according to Eq.~(\ref{quasigas}), where 
the summation over ${\bf P}$ is restricted to that region where bound states exist, i.e.\ $B^{\rm bind}_{A,\nu}({\bf P};T,n_B,Y_p) \ge 0$.

We denoted the density $n_{A,\nu}^{\rm Mott}(T,Y_p)$
as Mott density \cite{RMS,rop82,rop82a,ropsms} where the binding energy of a cluster $\{A,\nu\}$ with c.o.m. momentum ${\bf P}=0$ vanishes, 
\begin{equation}
\label{nMott}
 B^{\rm bind}_{A,\nu}(0;T,n_{A,\nu}^{\rm Mott},Y_p) =0\,,
\end{equation}
see Fig. \ref{Fig:Udo1}. For baryon densities $n_B>n_{A,\nu}^{\rm Mott}(T,Y_p)$ we introduced
the Mott momentum ${\bf P}_{A,\nu}^{\rm Mott}(T,n_B,Y_p)$, where the bound state disappears, 
\begin{equation}
\label{PMott}
B^{\rm bind}_{A,\nu}({\bf P}_{A,\nu}^{\rm Mott};T,n_{B},Y_p)=0\,.
\end{equation}
At $n_B>n_{A,\nu}^{\rm Mott}(T,Y_p)$,
the summation over the momentum to calculate the bound state contribution to the composition (\ref{eosq}) is restricted to the region 
$|{\bf P}| >|{\bf P}_{A, \nu}^{\rm Mott}(T,n_B,Y_p)|$. 

The condition (\ref{PMott}) may be replaced by further restrictions
in the $P$-space if further decay modes are considered.
Especially, the decay into $\alpha$ quasi-particles is of interest. 
This effect takes place, e.g., in the region $6 \le A \le 11$
where also the stability of the clusters with respect to
the decay into other fragments such as deuterons and tritons/$^3$He 
must be checked in order to determine $P^{\rm Mott}_{A \nu}$.

The Mott point where the binding energy vanishes is determined by the Pauli blocking term, 
$E_{A,\nu}^{0}+\Delta E_c^{\rm Pauli}({\bf P};T,n_{A,\nu}^{\rm Mott},Y_p)=0$.
The self-energies $\Delta E_{A,\nu}^{\rm SE}$  for the bound state and the continuum 
 states compensate if the momentum dependence is neglected.
Crossing the Mott point by increasing the baryon density, part of correlations survive as continuum correlations 
so that the properties change smoothly. 
Therefore, the inclusion of correlations in the continuum is of relevance.

\subsection{Excited states, continuum correlations and virial expansions}
The chemical equilibrium  in Eq. (\ref{eosq}) contains the sum over all components.
This includes not only the $A$-nucleon clusters in the ground state but also the $\nu$-summation over all excited cluster states.
This can be replaced by an integral after introducing the density
of states \cite{boh}
\begin{equation}
\label{38}
 D_A(E)=\frac{1}{12}\left(\frac{\pi^2}{a}\right)^{1/4} E^{-5/4} \exp (2 \sqrt{aE})
\end{equation}
with $a = A/ 15$ MeV$^{-1}$ for the homogeneous Fermi gas model
and arbitrary values $Z$. Then the abundance of the
clusters with mass number $A$ is given by
\begin{eqnarray}
\label{39}
&& \!\! \!\!  \!\!  X_A= \frac{n_A}{n_B}=\frac{A^3}{2 \pi^2 n_B \Lambda^3} \int_{P^{\rm Mott}_{A,0} \Lambda/A}^\infty x^2 dx
\int_{E_{\rm min}^A}^{E_{\rm max}^A} dE D_A(E)\\
&& \!\!  \!\!  \!\!  \!\!  \times \exp\left\{-\frac{1}{k_BT}\left[E^0_{A,0}(n_B)+E+\Delta E_{A,0}(n_B,x)
+A \frac{k_BT x^2}{4 \pi}-(A-Z) \mu_n-Z \mu_p \right] \right\}.\nonumber 
\end{eqnarray}
In this formula, it was assumed that Coulomb corrections,
self-energy and Pauli blocking corrections to the cluster
energy do not strongly depend on the excitation state. A
lower bond $E_{\rm min}^A+E_F/A$ ($E_F$ - nuclear matter Fermi
energy) was introduced to take into account that below
this energy the density of states (\ref{38}) is not applicable, and
the discrete structure of the excitation spectrum of the
clusters should be considered.
 This lower limit can be flexible depending on the number of states 
 which are explicitly taken into account.
 The upper bond $E_{\rm max}^A(x)$ is
introduced into (\ref{39}) to take into account that excited
clusters may become instable with respect to the decay
into smaller fragments 
as given, e.g., by $E_{\rm max}^A=E_{\rm Mott}^A=A \Delta E^{\rm SE}_\tau-E_{A,0}^0$ \cite{R81}.

The summation over all excited states $\nu$ is not restricted to the bound states but includes also scattering states.
Only taking into account the contribution of scattering states, the correct low-density limit of the EoS 
and related thermodynamic quantities is obtained. Expanding the EoS in powers of  $n_B$, 
the lowest  order gives the result for ideal quantum gases 
$n_B^{(0)}=2 \Lambda^3[\exp(\mu_n/T) +[\exp(\mu_p/T)]$ and for the pressure $p= n_BT$. 
The second order of $n_B$ is denoted as second virial coefficient.
An exact expression is given by the Beth-Uhlenbeck equation 
\begin{eqnarray}
\label{B-U}
&&n_{2,d}= \frac{2^{3/2}}{\Lambda^3} e^{\left(\mu_n+\mu_p\right)/T}  \left[3 e^{-E_d^0/T}+\int_0^\infty \frac{dE}{\pi} e^{-E/T}\frac{d}{d E} \delta_{2,d}^{\rm tot}(E) \right]
\end{eqnarray}
with
$\Lambda=(2 \pi \hbar^2/m T)^{1/2}$ being the baryon thermal wavelength (the neutron and proton masses are approximated by 
$m_\tau \approx m=939.17$ MeV$/c^2$), and $  \delta_{2,T_I=0}^{\rm tot}(E)= \sum_{S,L,J}(2J+1) \delta_{{^{2S+1}}L_J}(E) $ 
the isospin-singlet ($T_I=0$) scattering phase shifts with angular momentum $L$ 
as function of the energy $E$ of relative motion. A similar expression can also be derived for the isospin-triplet channel (e.g.\ two neutrons) 
where, however, no bound state occurs.  The relation (\ref{B-U}) gives an exact relation 
for the second virial coefficient in the low-density limit where in-medium effects are absent. For data see \cite{HS} 
where detailed numbers are given. Note that the second viral coefficient is expressed in terms of measured data, the binding energies and scattering phase shifts $\delta_2(E)$,

These second virial coefficients  cannot directly be used within a quasiparticle approach.
Because part of the interaction is already taken into account when introducing the quasi-particle energy, 
one has to subtract this contribution from the second virial coefficient to avoid double counting, 
see \cite{clustervirial,SRS}.
Instead of Eq. (\ref{B-U}) we obtain for the $d$ channel
\begin{eqnarray}
\label{n2virial}
&&n^{\rm qu}_{2,d}= \frac{2^{3/2}}{\Lambda^3} e^{\left(\mu_n+\mu_p\right)/T} \\ &&\times 
 \left[3 (e^{-E_d^0/T}-1)+\int_0^\infty \frac{dE}{\pi T} e^{-E/T}\left\{ \delta_{2,d}^{\rm tot}(E)-\frac{1}{2} \sin [2  \delta_{2,d}^{\rm tot}(E) \right]\right\}. \nonumber
\end{eqnarray}
Comparing (\ref{n2virial}) with the ordinary Beth-Uhlenbeck formula (\ref{B-U}) there are two differences:\\
i) After integration by parts, the derivative of the scattering phase shift is replaced by the phase shift, and according to the Levinson theorem 
for each bound state the contribution $-1$ appears. \\
ii) The contribution $-\frac{1}{2} \sin [2 \delta_c(E)]$ appears to avoid double counting \cite{SRS,clustervirial} when introducing 
the quasiparticle picture.

The EOS (\ref{eosq}) is not free of ambiguities with respect to the subdivision into bound state contributions and continuum contributions, compare
(\ref{n2virial}) and (\ref{B-U}).  
The continuum correlations in the second virial coefficient are reduced if the quasiparticle picture is introduced.
The remaining part of the continuum contribution 
 in Eq. (\ref{B-U}) is absorbed in the quasiparticle shift. 
This has been discussed in detail in \cite{SRS,clustervirial,VT}.

At higher densities, we can introduce also the quasiparticle picture for the $A=2$ channel so that the deuteron energy $E^0_d$ is replaced by the in-medium (quasiparticle) deuteron energy, and the phase shifts $\delta_c(E)$ contain also the medium modifications, see \cite{SRS}. The approximation based on the solution of the in-medium two-particle problem (\ref{wave2})
leading to the generalized Beth-Uhlenbeck formula is denoted as (3, medium) in Tab. 2.

\subsection{Cluster virial approach and correlated medium}

A more advanced approach  [(4, medium) in Tab. 2] to the nuclear matter EoS would include cluster with arbitrary $A$.
A cluster Beth-Uhlenbeck approach is discussed recently \cite{clustervirial} to include also higher-order ($A>2$)
 correlations \cite{R15}.

For the $A$-nucleon cluster, the  in-medium Schr\"odinger equation (\ref{waveA}) is derived, depending on the occupation numbers 
$n(1;T,\mu_n,\mu_p)$ of the single-nucleon states $|1\rangle$. 
It is obvious that the nucleons found in clusters contribute to the mean field leading to the self-energy, 
but occupy also phase space and contribute to the Pauli blocking. 
The cluster mean-field (CMF) approximation \cite{clustervirial,R15,ropsms} considers also the few-body t-matrices in the self-energy 
and in the kernel of the Bethe-Salpeter equation. The CMF approximation leads to similar expressions (\ref{3}) 
but the free-nucleon Fermi distribution
$f_{1,\tau_1}(1)$ replaced by the effective occupation number (\ref{effocc}).

Because the self-consistent determination of $n(1;T,\mu_n,\mu_p)$ for given $\{T, \mu_n, \mu_p\}$ is very cumbersome,
as approximation the Fermi distribution with new
parameters $\{T_{\rm eff},\mu_n^{\rm eff}, \mu_p^{\rm eff}\}$ (effective temperature and chemical potentials) have been introduced  \cite{R15},
\begin{equation}
\label{effparameter}
n(1;T,\mu_n,\mu_p)\approx f_{1,\tau_1}(1; T_{\rm eff},\mu_n^{\rm eff}, \mu_p^{\rm eff})\,.
\end{equation}
The effective chemical potentials $\mu_\tau^{\rm eff}$ realize the normalization to the given baryon densities $n_\tau^{\rm tot}$.
A simple relation
\begin{equation}
\label{Teff}
 T_{\rm eff} \approx 5.5\, {\rm MeV} + 0.5\,\, T + 60\,\, n_B\,\, {\rm MeV\,\,fm}^3
\end{equation}
was given in Ref. \cite{R15} as an approximation for the region 5 MeV $< T <$ 15 MeV and densities $n_B < n_{\rm sat}/2$
of the parameter space. More detailed investigations are necessary to derive a more general expression
for the effective temperature as function of $T,n_B,Y_p$ including, for instance $\alpha$ matter where the medium consists of $\alpha$ nuclei. The present simple fit formula (\ref{Teff}) may be considered as a first step
in this direction.

\subsection{Coulomb correlations and Debye-Thomas-Fermi
screening}\label{Sec:2.1}

In Refs. \cite{RMS,Debrecen,R83} the Debye-Thomas-Fermi theory 
known from plasma physics was adapted to
calculate the Coulomb corrections to the cluster energies
 $E^0_{A \nu P}(T,n_B,Y_p)$ and the nucleon density variations (pair distribution function) within the
screening cloud around a cluster. A more sophisticated
approach to these Coulomb corrections can be formulated
by considering the cluster self-energy due to the Coulomb
part of the interaction in the dynamically screened
potential approximation, see, for instance, Refs. \cite{ebe,KKER}.

The screened potential $V_D ( r)$ is obtained from the Poisson
equation, and the screened density follows from the self-consistent solution of the linearized Poisson-Boltzmann
equation. A typical parameter is the screening length $1/\kappa$ where
\begin{eqnarray}
\label{18}
 \kappa^2=2 \pi e^2\frac{\partial}{\partial \mu_p}n_p(\beta,\mu_n,\mu_p).
\end{eqnarray}

Two effects are obtained from the Coulomb interaction:

(i) The density  of the surrounding charged particles is reduced in the vicinity of the cluster.
The total proton density $n_p^*.$ at the surface of a cluster is smaller than the average density $n_p^{\rm tot}$
because of the Coulomb repulsion.\\
This effect will diminish the in-medium corrections due
to the short range nucleon-nucleon interaction especially
for large clusters.

(ii) In addition to the Coulomb energy of the isolated nucleus,
i. e. $n_B = 0$, a self-energy shift due to the finite charged nucleon
density $n_p^{\rm tot}$ is obtained. For  globally charge-neutral matter, in particular in high-density astrophysical objects,
the Coulomb term in the Bethe-Weizs\"acker 
formula is reduced. The Coulomb field of the charged nucleus extends for  $n_B = 0$ over the entire space, 
but is confined to the compensating screening cloud at finite matter density. 
This means a reduction of the electric field energy. To give an estimate, the simple 
Debye theory gives shift $-Z^2e^2\kappa/2$.\\
This effect makes the large clusters more stable. 
With increasing density, the valley of stable nuclei is shifted towards the symmetry line $Z=N$. Larger clusters can be 
formed because the destabilizing influence of the Coulomb term is reduced.

Expressions to calculate both effects (i) and (ii) are given in the literature \cite{RMS,Debrecen,R83}.

Note that a simplified description of the effects of Coulomb
correlations is given by the Wigner-Seitz cell method
as used, for instance, in Ref.
\cite{bar}. In the spirit of the Wigner-Seitz model, all
protons are removed from a sphere with the radius
$R^{\rm WS}_A = (3 Z/4\pi n_p^{\rm tot})^{1/3}$ so that $n_{p}^{ {\rm WS}}( r)=n_p^* = 0, r < R^{\rm WS}$, and
$ n_{p}^{{\rm WS}}(r) =n_p^{\rm tot}$ elsewhere. 
Instead of the Debye shift which reads in the classical limit (uncorrelated medium)
\begin{equation}
 \Delta E^{\rm Coul,D}_{A \nu P}(T,n_B,Y_p)=-\frac{Z^2 e^2}{2} \left( \frac{4 \pi e^2 n_p^{\rm tot}}{m T}\right)^{1/2}
\end{equation}
we have ($R_A=1.25 A^{1/3}=(3/4 \pi n_{\rm sat})^{1/3}$ fm)
\begin{equation}
 \Delta E^{\rm Coul,WS}_{A \nu P}(T,n_B,Y_p)=-\frac{3}{5}\frac{Z^2 e^2}{ R_A} \left[ \frac{3}{2}\frac{R_A}{R^{\rm WS}_A}-
 \frac{1}{2}\left(\frac{R_A}{R^{\rm WS}_A}\right)^3 \right].
\end{equation}

Light clusters $A \le 4$ are not significantly influenced by Coulomb interaction.
In contrast,
Coulomb interaction is of fundamental interest for the stability of large clusters,
and it is very important in the phase transition region determining the pasta-like structures.
The simple Wigner-Seitz model can give only a rough estimate of the effects of Coulomb interaction.
It has to be replaced by more accurate calculations
which treat the Coulomb correlations within a QS approach.  
The solution of the Poisson-Boltzmann equation with the consistent account of normalization 
gives an adequate treatment of Coulomb effects.

\section{Experimental evidences and relevance}

We discuss several properties which can be observed. We give not an exhaustive description of these properties,
but intend only to point out the relevance of clustering in nuclear systems and the necessity to describe medium effects.
 \begin{figure}[ht]
 			\includegraphics[width=0.8\linewidth,clip=true]{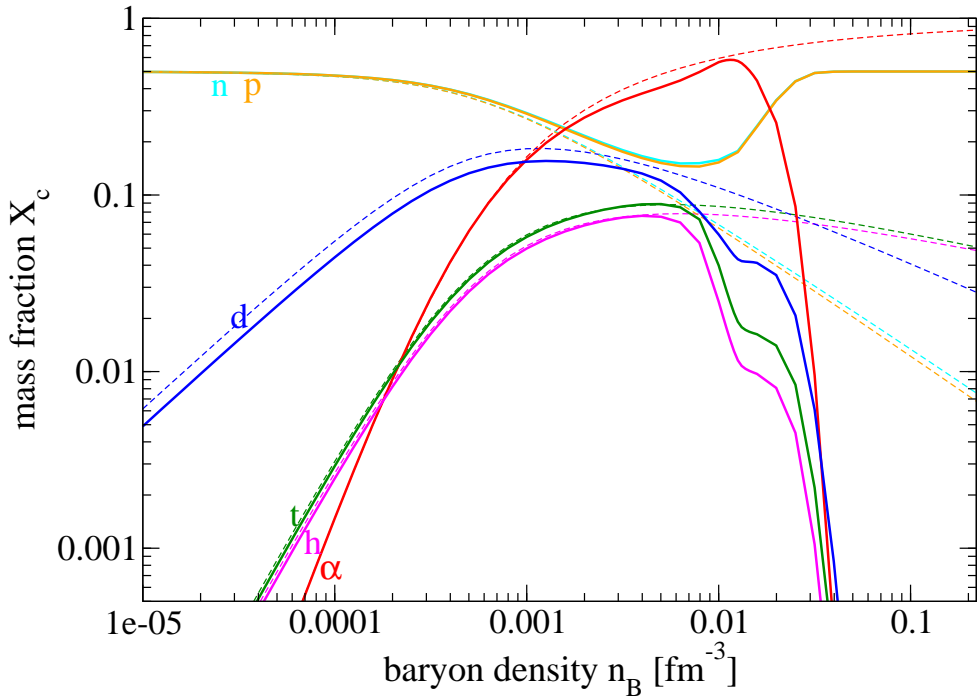}
 			\caption{Composition of symmetric matter in dependence on the baryon density $n_B$,
			$T=5$ MeV. Quantum statistical calculation \cite{R15} compared with NSE.
 }
 			\label{Fig:composition}
 		\end{figure}

\subsection{HIC: EoS and chemical constants}

Quantum statistical (QS) calculations, Eq. (\ref{eosq}), of the composition of symmetric ($Y_p=0.5$) matter 
are shown in Fig. \ref{Fig:composition}  \cite{R15}. Deviations from the simple NSE at low densities are caused by the account of scattering states (in particular $d$). For densities $n_B > 0.001$ fm$^{-3}$ medium effects become of relevance. Pauli blocking leads to the disappearance of bound states around $n_B=0.03$ fm$^{-3}$ so that a Fermi liquid of single nucleon quasiparticles remains.

The experimental verification of clustering in nuclear systems is not simple.
First, in laboratory experiments we have not infinite nuclear matter, but always finite systems.
Second, experiments like heavy ion collisions (HIC) are non-equilibrium processes
and are only approximately described by equilibrium properties, 
for instance within a freeze-out approach. Despite the consistent quantum statistical 
description of cluster formation in HIC may be solved by future transport codes,
the local thermodynamic equilibrium is a prerequisite for such a nonequilibrium approach,
not only as a test case for the quality of the equilibrium limit solution,
but also for the formulation of the kinetic theory beyond the Boltzmann single-nucleon description.

The cluster yields and respective energy spectra have been discussed intensely to derive 
the thermodynamic properties of nuclear matter at finite temperature. 
Recently investigations have been published \cite{Natowitz} which clearly rule out the NSE, 
but demonstrate the relevance of medium corrections. Within a QS  approach
including quasiparticle shifts and correlations in the continuum \cite{R15}, 
it was possible to reproduce the data for the chemical constants
of the light elements $d, t, h, \alpha$ obtained from the cluster yields, see Fig \ref{Fig:composition}. More simple 
models for medium corrections such as the excluded volume concept \cite{Hempel} 
can be adapted to reproduce the data \cite{Hempel2014}.

So-called Mott points have also been discussed when comparing the theory of cluster formation in dense
matter with experiments \cite{HagelMott}. Note that for $n_B>n^{\rm Mott}_{A,\nu}$ the abundance 
of the cluster $\{A,\nu\}$ is not vanishing. 
Contributions from high momenta ${\bf P}>P^{\rm Mott}_{A,\nu}(T,n_B,Y_p)$ and the continuum remain.

\subsection{EoS and gas-liquid nuclear matter phase transition}

\begin{figure}[ht]
 			\includegraphics[width=0.8\linewidth,clip=true]{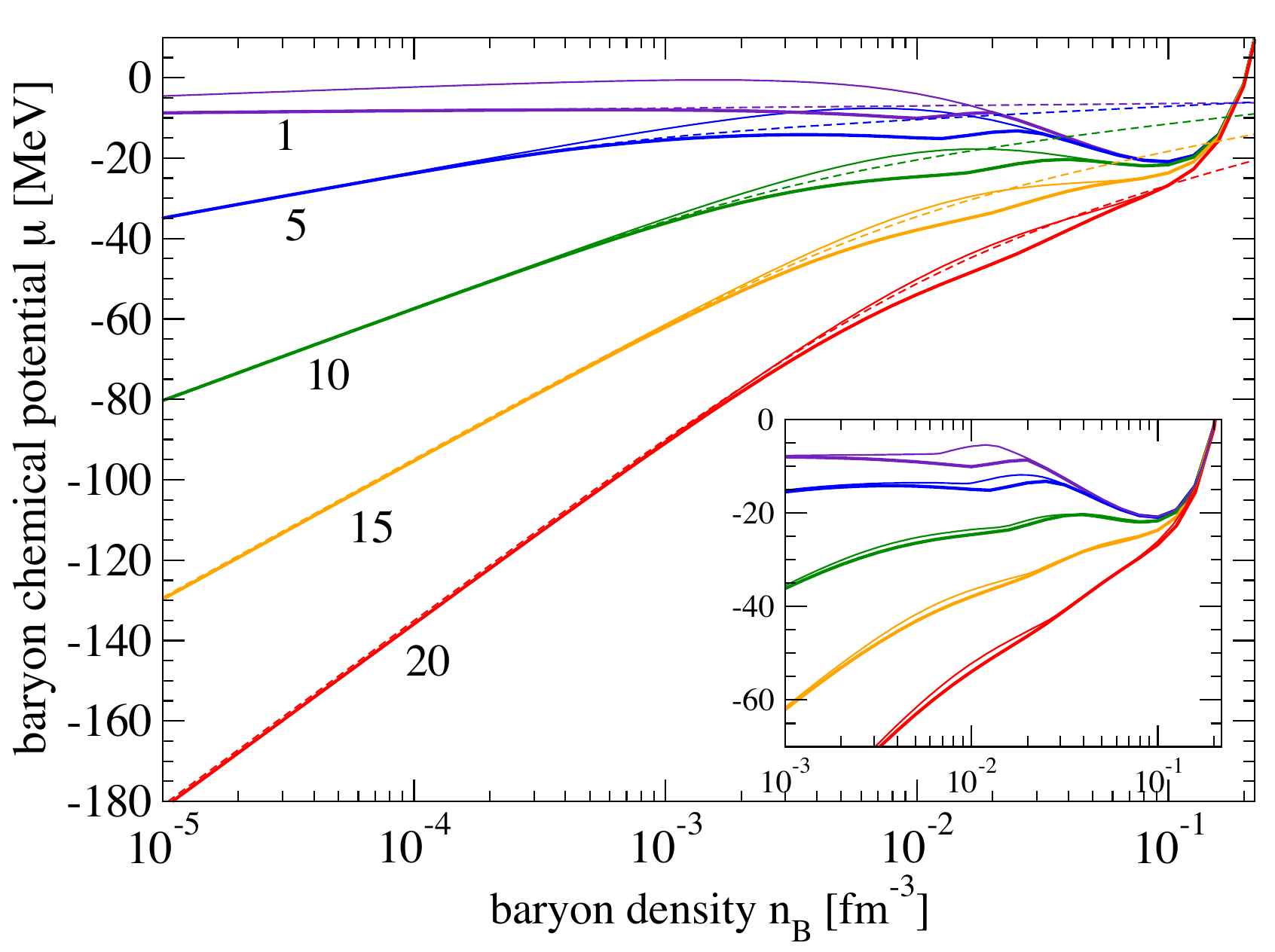}
 			\caption{Chemical potential for symmetric matter. QS calculation compared with RMF (thin) and NSE (dashed). Insert: QS calculation without continuum correlations (thin lines). From \cite{R15}.
 }
 			\label{Fig:mu}
 		\end{figure}

The evaluation of the EoS (\ref{eosq}) for symmetric matter ($Y_p=0.5$) is shown in Fig. \ref{Fig:mu} for different $T$.
Further EoS such as the Free energy $F(T,N_n,N_p,\Omega)$ are obtained by integration; see Sec. \ref{Sec:1.1}.
We will not present results for all thermodynamic quantities, see Ref. \cite{Typel}, but discuss only the influence of clustering.
RMF is applicable at high densities where clusters disappear. The ideal Fermi gas, valid at very low densities, is improved by
the NSE which becomes invalid around $n_B=0.001$ fm$^{-3}$.

An interesting point is the stability of the EoS for homogeneous matter with respect to phase transitions.
For instance, if $\partial \mu/\partial n_{|_T} < 0$, separation of phases with different densities will occur.
 We would like to point out that the parameter values of
 $T$ and $n_B$ considered here lie within that region of the 
 temperature-density plane of matter where a nuclear matter
 phase transition is possible \cite{kup,bar,lam,fri,rop82,rop82a,sch82}. 
 Of interest is the reduction of the region of phase instability obtained from RMF, 
 if correlations and cluster formation are taken into account; see Fig.  \ref{Fig:mu}.
 The critical temperature $T^{\rm mf}_{\rm cr}=13.72$ MeV, see Sec. \ref{Sec:2.1}, 
 is reduced to $T^{\rm QS}_{\rm cr}=12.42$ MeV \cite{R15,rop82,ropsms}, se also \cite{shenTeige,Vergleich}.
 
The spinodal instability has been considered \cite{Udo,Udo1} and a limiting temperature 
for large clusters has been discussed as limit for thermalization owing to spinodal vaporization.
This  may be improved by considering the intrinsic partition function, in particular the continuum contributions. 
For homogeneous matter, phase separation is suppressed because of the long-range Coulomb interaction.
Structures (droplets, wires, sheets, etc.) are formed separating high-density regions from low-density regions. 
Such so-called pasta phases give lower values for the free energy.
 So-called nuclear pasta phases which may have complex structures, 
 are discussed to derive the EOS also within the region of thermodynamic instability, 
 see \cite{Furusawa,Avancini,Vergleich,Constanca}.

 Note that cluster formation
 in hot dense matter has also been treated as phase separation and droplet formation
 near the phase transition region. In Ref.  \cite{lam},
 the dense phase is assumed to be distributed
 as droplets (identical nuclei) immersed in the low dense
 phase which consists of free nucleons and ideal $\alpha$ particles. 
 Within a finite temperature Thomas-Fermi
 approach, abundances of clusters and probabilities
 of fluctuations of the droplet size were considered  \cite{bar,buc}.
 A mass formula approach which takes into account excited
 nuclei and Coulomb effects was given in Ref. \cite{maz}. These approaches,
 however, do not seem to be adequate to describe the
 abundance of small clusters where a more rigorous quantum
 statistical approach is necessary. Confusion arises when a
 nucleus which is a bound quantum state is treated as
 droplet of a dense phase which is a classical object.

\subsection{Symmetry energy}
 \begin{figure}[ht]
 			\includegraphics[width=0.7\linewidth,clip=true]{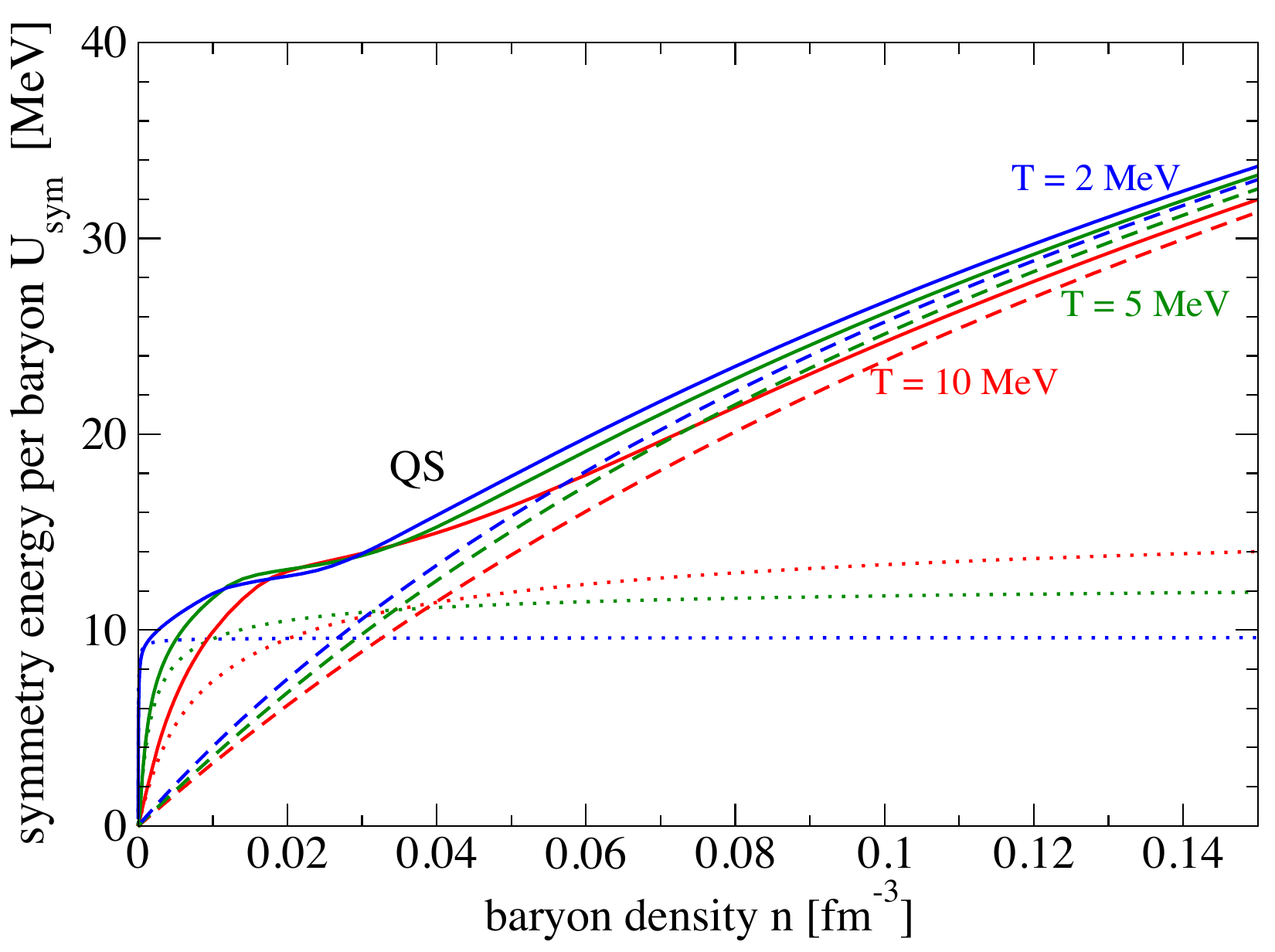}
 			\caption{Symmetry energy for different $T$. QS calculation compared with RMF (dashed) and NSE (dotted). From \cite{Hagel}.
 }
 			\label{Fig:sym}
 		\end{figure}

Solving the EoS (\ref{eosq}), after calculating the free energy as discussed above
further thermodynamic quantities are obtained such as the internal energy per nucleon $U(T,n_B,Y_p)$.
The symmetry energy $U_{\rm sym}(T,n_B)$ describes the dependence of the internal energy
on the asymmetry parameter $Y_p$ and is related to the difference 
of the internal energy of neutron matter and symmetric matter, see \cite{Li,Hagel}.

The symmetry energy is sensitive to cluster formation, see Fig. \ref{Fig:sym}.
Whereas quasiparticle approaches such as Skyrme Hartree-
Fock and relativistic mean-field (RMF) models predict in the low-density limit
$U_{\rm sym}(T,n_B) \propto n_B$ \cite{Li}, the QS calculations \cite{R15,Hagel}
show strong deviations at low densities compatible with the NSE. Cluster formation
is strongly $T$ dependent, and the low-density, low-temperature limit is dominated 
by the binding energy per nucleon in nuclei which is $\approx 8$ MeV.

Such a finite value of $U_{\rm sym}(T,n_B)$ in the low-density region in 
contradiction to the mean-field approaches has been confirmed experimentally
by \cite{Kowalski,Wada,Natowitz}. At low density the symmetry energy changes mainly
because additional binding is gained in symmetric matter
due to formation of clusters and pasta structures \cite{Wata}.

\subsection{Low temperatures and quantum condensates}

A special feature of correlations and bound state formation are quantum condensates.
According to Eqs. (\ref{vert}), (\ref{eosq}), the Bose distribution function exhibits a singularity 
when the energy eigenvalue $E_{A \nu}({\bf P};T,\mu_n,\mu_p)$ coincides with the cluster chemical potential
$\mu_A=(A-Z) \mu_n+Z \mu_p$ (Thouless criterion). The case $A=2$ is well investigated.
Whereas in the low-density limit, depending on the asymmetry, Bose-Einstein condensation (BEC) of
deuterons is expected to occur, at high densities Bardeen-Schrieffer-Cooper pairing (BCS) of continuum states happens. The dissolution of bound states 
is connected with the crossover from BEC to BCS \cite{Almpn}. These effects are described 
solving the two-nucleon wave equation (\ref{wave2}).

However, in symmetric matter the formation of a BEC of deuterons interferes with the BEC of $\alpha$ particles (quartetting)
which are strongly bound (7.1 MeV/$A$ for $\alpha$ in contrast to 1.1 MeV/$A$ for $d$) so that, at finite temperature, with increasing chemical potential
(increasing density) the BEC of $\alpha$ particles occurs prior to the BEC of deuterons \cite{RSSN}. The BEC  of $\alpha$ particles
disappears abruptly when the bound states are dissolved owing to the Pauli blocking, which follows from the solution
of the in-medium wave equation (\ref{waveA}) for the special case $A=4$, $E_\alpha=\mu_4$.

Quantum condensates are not rigorously incorporated in present standard EoS. 
Whereas pairing is rather well described, the transition 
from $\alpha$ matter to nuclear matter at low temperatures is not clearly described until now. 
Experimentally, signatures of pairing (rigorously defined for infinite matter) are seen in the 
even-odd staggering of the binding energies of nuclei. Signatures of quartetting have been identified for the Hoyle 
state of $^{12}$C \cite{THSR,Fun08},  see the following section \ref{Sec:nucStruc}.

\subsection{Nuclear structure}
\label{Sec:nucStruc}

Results derived for homogeneous nuclear matter cannot directly applied to  finite nuclei. 
Only within a density functional  approach, the local density approximation uses these results.
Neglecting any correlations, the mean-field approximation [(1, medium) of Tab.~1] has been applied very successful
as shell model of nuclei. The many-nucleon wave function is approximated by the antisymmetrized product
of single-nucleon states which are not momentum eigenstates as in homogeneous matter 
but have to be determined self-consistently.

Inclusion of correlations is possible, for instance, by superposition of shell-model states which needs some effort to
realize a cluster structure. Other approaches implement the formation of clusters from the beginning,
for instance the resonating group method and related approaches, for a review see \cite{REVTHSR}. 
Numerical simulations such as fermionic/antisymmetrized molecular dynamics (FMD/AMD) are at present restricted to
small numbers ($A\le 12$) of nucleons.

There is clear evidence for clustering in nuclei if the density is low. 
We give some examples for $\alpha$ clustering in nuclei.  
Symmetric 4$n$ nuclei ($^8$Be, $^{12}$C, $^{16}$O, etc) in dilute gas states near the $n \alpha$ threshold have been investigated, experimentally and by theory. 
A famous example is the Hoyle state which is an excited state of $^{12}$C, see Refs. \cite{THSR,REVTHSR}.
Antisymmetrization of the total wave function (Pauli blocking) is responsible that in contrast to the low-density Hoyle state
(large rms radius $\approx 4.29$ fm) where $\alpha$-like clusters are well established, in the dense ground state $^{12}$C
(rms radius $\approx 2.65$ fm) the uncorrelated product state (Slater determinant) is a good approximation.
Clustering is also visible in low-lying breathing excitations of nuclei as well as in nuclear reactions. 
Another signature of quartetting my be the Wigner contribution \cite{Schnell} to the binding energy of near $Z=N$
nuclei.

A long-standing issue of correlations in nuclei is the $\alpha$ decay of heavy nuclei where a preformation of the 
$\alpha$ particle is assumed, see Ref. \cite{Po,Xu,Yokohama14} and further references given there. 
 $\alpha$-like correlations appear in the low-density tails at the surface of heavy nuclei which are $\alpha$ emitters. 
 For example, in $^{212}$Po a quartet $\{n_\uparrow,n_\downarrow,p_\uparrow,p_\downarrow\}$ 
 moves on top of the double-magic $^{208}$Pb core nucleus. Outside a critical radius $r_{\rm cr} = 7.44$ fm
 the nucleon density of the core nucleus is small, $n_B < 0.03$ fm$^{-3}$, so that an $\alpha$-like bound state 
 can be formed, whereas inside the core nucleus the quartet is described by uncorrelated single-nucleon states.
Improving the local density approach, a rigorous description of the quartet state moving on top of a core nucleus has to be worked out in future.

\subsection{Astrophysics}

An important application for the nuclear matter EoS is the structure and 
evolution of compact objects. Simulations of supernova explosions are performed recently
to compare with observed signals. The thermodynamic parameters 
of the scenario of core collapse, Fig. \ref{Fig:SN}, are found in the warm dense matter regime. 
Cluster formation is relevant to reproduce a consistent EoS \cite{SR}. 

 In particular, the physics of core collapse supernovae enters the parameter region where cluster formation
 with $A\leq 4$ in the subsaturation region occurs \cite{Tobias}. 
 The presence of clusters modifies the thermodynamic properties and affects, for instance, the neutrino transport
  \cite{Arcones2008,Furusawa,Connor,neutrinoS}. Whereas previous approaches \cite{LS,Shen} considered only $\alpha$ particle 
 formation, recently also other light elements have been taken into account, within a quantum statistical model \cite{SR} 
 or using the excluded volume concept \cite{Hempel}. 

A detailed knowledge about the supernova explosion
including the neutrino transport is necessary to answer different questions such as the emission of matter by explosions
or the cooling rates of pre-neutron stars, which is influenced by cluster formation and the occurrence of quantum phases.
The formation of heavy elements is an essential, unsolved problem in astrophysics. It needs a hot, neutron-reach environment.
At present it remains unclear to determine the site, for instance supernova explosions or  neutron star mergers, where the $r$-process occurs \cite{Arcones?,Kor12}.

Another topic is the structure of neutron stars. Parameter values $\{T,n_B,Y_p\}$ appear in the crust  where clusters formation occurs.  
Heavy nuclei are immersed in an environment consisting mainly of neutrons and electrons. At increasing density,
pasta phases are formed. The crust/core transition is presently discussed \cite{Constanca}, in particular the existence of a first-order phase transition. Inside the core, clusters (nuclei) are dissolved because of Pauli blocking.

\section{Results and Discussion}

Based on a quasiparticle concept, the present work aims at deriving the EoS for warm dense matter in the subsaturation region, 
incorporating the known low-density virial expansions as well as mean-field theories near saturation density. 
Different ingredients have been used: \\
(i) A cluster-virial expansion describes not only the different bound state contribution to the EoS, like the NSE, 
but takes into account also the contribution of the continuum. 
For $A=2$, according to the Beth-Uhlenbeck formula the contribution of the continuum is given by the two-nucleon scattering 
phase shifts. Introducing single-nucleon quasiparticle energies, double counting of the mean-field terms has to be avoided.\\
(ii) Medium modified bound state energies and scattering phase shifts are used. 
They result from the solution of a few-nucleon wave equation which contains mean-field single-nucleon energy shifts as 
well as Pauli blocking terms. Both should be calculated taking into account self-consistently correlations and bound state formation in the medium.\\
(iii) In homogeneous (stellar) matter, screening of the Coulomb interaction contributes to the medium modification 
of quasiparticle energies, in particular for large clusters, and pasta-like structures in the phase transition region.

Whereas for charged particle systems the Coulomb interaction is exactly known and a systematic quantum statistical treatment is well elaborated, 
see, e.g., \cite{KKER}, a fundamental $N-N$ interaction is not at our disposal. 
Nevertheless, the quasiparticle properties are well defined from correlation functions which can, in principle, be measured.
A semi-empirical approach has been used to calculate the quasiparticle properties after introducing an effective $N-N$ interaction 
adjusted to known properties of nuclear systems.

In the zero-density limit, we can avoid the solution of the $A$-nucleon wave equation (\ref{waveA}) using the empirical energies 
$E^0_{A \nu P}$ \cite{Audi} to evaluate the EoS (\ref{eosq}) resulting in the NSE. 
The second virial coefficient is determined by the measured scattering phase shifts.

Similarly, empirical data for properties near the saturation density are used to parametrize 
the single-nucleon quasiparticle energy shifts $E_\tau({\bf p}; T,n_B,Y_p)$ by a Skyrme force or RMF expressions.
The same is also possible for the quasiparticle energy shift $E_{A \nu P}(T,n_B,Y_p)$ 
of the light clusters $d,t,h, \alpha$
where empirical values for the rms radius or more details about the wave function can be used \cite{R2011}. 
The contribution of continuum correlations as well as a correlated medium is discussed in \cite{R15} 
for the light elements $A \le 4$.
A microscopic approach for the quasiparticle energy shifts solving the Brueckner equations for the single-nucleon case
or the in-medium Schr\"odinger equation (\ref{waveA}) for the $A$-nucleon case would be of interest for future work,
but demands an expression for the $N-N$ interaction. At present, only the case $A=2$ has been treated this way \cite{SRS,Urban}.
 Evidence for clustering at low densities and medium modifications are obtained from nuclear structure investigations. 

Less investigations have been performed for the light metals $5 \le A \le 11$. Because their binding energies are weak 
($^8$Be is unbound) and strongly influenced by the medium, see \cite{Debrecen}, their abundances are strongly reduced
if comparing with the NSE, see results of 
HIC experiments \cite{andro} but also calculations for stellar matter \cite{Rabund,Rop83} and for cosmic rays \cite{WZbla,cosmicrays}. 

Comparing with the light elements, the internal structure of heavy elements $A \ge 12$ (including excited states) is not drastically influenced by medium effects. The interaction with the surrounding nucleons is determined by Pauli blocking which 
is reflected by the concept of excluded volume. For heavy nuclei, an upper limit for the account of excited states has been introduced to get convergent results at higher temperatures. 

In addition to the strong interaction also the Coulomb interaction has to be treated, see Sec. 2.6. In particular, it is of relevance for large clusters and for pasta structures in the region of phase instability. 
Future investigations are necessary the include heavy elements as well as
pasta structure formation, especially in the region which is characterized by the thermodynamic instability of symmetric nuclear matter. 
Another challenging issue for a general EoS is the account of quantum condensates (pairing, quartetting) at low temperatures.

\section{Outlook}

 For the interpretation of HIC results, the thermodynamic relations such as the EoS
 describing infinite systems in thermodynamic equilibrium are not directly connected with the measured cluster yields and their 
 energy spectra because the nuclear system is inhomogeneous in space and, because of strong 
 non-equilibrium, inhomogeneous in time. An adequate description should consider kinetic equations 
 for the distribution functions (Wigner functions) of all clusters which have as equilibrium solutions not the 
 ideal Fermi gas but an appropriate approximation of the EoS, see Tab. \ref{Tab:2}. Thus, the EoS containing 
 quasiparticle clusters (medium-modified nuclei) may be considered as a prerequisite 
 to formulate a transport code for the nonequilibrium evolution. This is
described by the extended von Neumann equation for the statistical operator 
$\varrho(t)=\lim_{\varepsilon \to 0}\varrho_{\varepsilon}(t)$ \cite{ZMR}, 
\begin{equation}
 \frac{\partial}{\partial t}\varrho_{\varepsilon}(t)+\frac{i}{\hbar}\left[H,\varrho_{\varepsilon}(t)\right]
=- \varepsilon \left(\varrho_{\varepsilon}(t)-\varrho_{\rm rel}(t)\right).
\end{equation}
 The relevant statistical operator $\varrho_{\rm rel}(t)$ is obtained from the maximum of entropy 
 reproducing the local, time dependent composition with parameter values $T({\bf r}, t), \mu_n ({\bf r}, t),\mu_p({\bf r}, t)$, but contains in addition the cluster distribution functions $f^{\rm Wigner}_{A\nu}({\bf p,r},t)$ as relevant observables \cite{Zub93,BUU}.
 
 Even if we can define a freeze-out state (temperature and chemical potentials) which determines the main features of 
 the composition, further reaction and decay processes will occur before the cluster yields, observed in the detectors, are established.
 In this context it is of interest not only the decay of excited and unbound (e.g. $^8$Be) nuclei, 
but what happens with the continuum correlations which are present at high densities. 
For instance, in the $n-n$ channel where no bound state arises, all continuum correlations contribute to the neutron distribution function. 
 In contrast, in the $n-p$ channel part of the continuum correlations contributes to the deuteron distribution, 
 whereas the remaining part is found in the distribution of free neutrons and protons; for a discussion see Ref. \cite{R15}. 
 Future work is necessary 
 to devise a transport theory for HIC
 which is compatible with the thermodynamic properties and the EoS, described in this work, as equilibrium solution \cite{Pawel,Csernai,Beyer}.
 
 In this context it is also of interest to find optimum parameter sets $\{T, n_B,Y_p\}$ 
 for the reproduction of observed abundances of clusters. 
This has been done for HIC where the yields of light clusters have been used to infer 
the thermodynamic parameter values \cite{Natowitz,densitometer}. 
In contrast to a simple chemical equilibrium such as the Albergo thermometer or densitometer which is connected with the ideal mixture of different components (NSE), density effects are of relevance. The freeze-out parameter represent a state during the evolution where 
local thermodynamic equilibrium ceases to be realized approximately. 
The further evolution is characterized by the different cluster distribution functions. 
It is determined by collisions, reactions, and decay processes. 
 
This description can also be applied to astrophysical abundances of elements.  Only the gross properties of elemental distribution are described by a freeze-out approach. Details are related to further reactions during cooling and expansion, forming local (with respect to the $N-Z$ plane) deviations. Based on the  cluster distribution functions $f^{\rm Wigner}_{A\nu}({\bf p,r},t)$ as relevant observables, a reaction network can describe this stage of evolution of the nuclear system.
 
As an example, we can compare solar element abundances $X_A = n_A / n_B$ \cite{cam} with calculations of the abundances 
 for hot dense matter. Reasonable agreement with the gross behavior of the solar abundances was obtained  with
 parameter values temperature $k_B T = 5$ MeVand nucleon density
  $n_B = 0.016$ fm$^{-3}$ 
  \cite{Rabund,R83,Rop83,Debrecen}. 
Other parameter values, for instance $k_B T = 5$ MeV, nucleon density
 $n_B = 0.0156$ fm$^{-3}$ as well as $k_B T = 5.5$ MeV, nucleon density
 $n_B = 0.0168$ fm$^{-3}$ can be associated with the chemical composition of different stars \cite{WZbla}. 
 Cosmic ray abundances are parametrized  with a higher  temperature $k_B T = 5.88$ MeV, nucleon density
 $n_B = 0.018$ fm$^{-3}$\cite{cosmicrays}.   The asymmetry variable $Y_p=0.5$ was not optimized.

 Medium modifications are of relevance for these parameter values.
   In addition to the deviations from the NSE for light elements, we have a strong depletion due to Pauli blocking for weakly bound light metals $5\le A \le 11$. Note that the origin of elements is not completely solved until now. In particular,
the site where heavy elements are produced, is not identified yet. In this
connection it is of interest to ask for parameter sets $\{T , n_B, Y_p \}$ which optimally reproduce the observed abundances.

\section*{Acknowledgements}

The author is thankful for discussions with D. Blaschke, M. Hempel, J. Natowitz, W.U. Schroeder, 
A. Sedrakian, P. Schuck, S. Typel, and H.H. Wolter.

\end{document}